 \documentclass[aps,prc,showpacs,floatfix,4paper]{revtex4}
\usepackage{amsmath}
\usepackage{amssymb}
\usepackage{epsfig}

\newcommand{\be}{\begin{eqnarray}}
\newcommand{\ee}{\end{eqnarray}}

\def\bfg #1{{\mbox{\boldmath $#1$}}}

\begin{document}
\title{Forward $\bar p d$ elastic scattering and total spin-dependent
$\bar p d$ cross sections at intermediate energies} 

\author{Yu.N.~Uzikov$^1$, J.~Haidenbauer$^2$}

\affiliation{
$^1$Laboratory of Nuclear Problems, Joint Institute for Nuclear
Research, 141980 Dubna, Russia\\
$^2$Institut f\"ur Kernphysik and J\"ulich Center for Hadron Physics,
Forschungszentrum J\"ulich, D-52425 J\"ulich, Germany
}

\begin{abstract}
Spin-dependent total ${\bar p}d$ cross sections 
are considered using the optical theorem. For this aim 
the full spin dependence of the forward 
${\bar p}d$ elastic scattering amplitude is considered in a model 
independent way. The single-scattering approximation is used
to relate this amplitude to the elementary 
amplitudes of $\bar p p$ and $\bar p n$ scattering 
and the deuteron formfactor.
A formalism allowing to take into account Coulomb-nuclear
interference effects in polarized $\bar p d$
cross sections is developed.
Numerical calculations for the polarized total 
${\bar p}d$ cross sections are performed at beam energies 
20-300 MeV using the $\bar N N$ interaction models developed by
the J\"ulich group.
Double-scattering effects are estimated within the Glauber approach
and found to be in the order of 10 - 20 \%.
Existing experimental data on differential $\bar p d$ cross sections 
are in good agreement with the performed Glauber calculations.
It is found that for the used ${\bar N}N$ models  
the total longitudinal and transversal ${\bar p }d$
cross sections are comparable in absolute value to those for
${\bar p}p$ scattering. 
\end{abstract}

\maketitle

\section{Introduction}
Recently the PAX collaboration was formed \cite{PAX} with the aim
to measure the proton transversity in the interaction of polarized 
antiprotons with protons at the future FAIR facility in Darmstadt. 
In order to produce an intense beam of polarized antiprotons, the collaboration 
is going to use antiproton elastic scattering off a polarized hydrogen 
target ($^1$H) in a storage ring \cite{Rathmann}.
The basic idea
is connected to the result of the FILTEX experiment \cite{FILTEX},
where a sizeable effect of polarization buildup was achieved in a 
storage ring by scattering of 
unpolarized protons off a polarized hydrogen atoms at low
beam energies of 23 MeV. 
 
According to recent theoretical analyses \cite{MS,NNNP,NNNP1,NNNP2}
the polarization effect observed in Ref. \cite{FILTEX} has to be interpreted
in such a way that solely the spin dependence of the hadronic (proton-proton) 
interaction provides the spin-filtering mechanism, i.e. is responsible for
different rates of removal of protons from the ring for different initial 
polarization states.
In other words \cite{MS,NNNP},
proton scattering on the polarized electrons of hydrogen
atoms cannot provide a sizeable effect of polarization buildup, as it was
assumed before \cite {HOM}. Indeed, the maximal scattering angle in
this process, $\theta_{max}=m_e/m_p=0.5$ mrad, is less than the beam
acceptance angle $\theta_{acc}$, which is defined so that for scattering at 
smaller angle $\theta < \theta_{acc}$ the
projectile remains in the beam.
 For this case, the beam-into-beam scattering
kinematics of this process in a storage ring allows the proton polarization 
buildup only due to spin-flip transitions between the initial and final 
spin states of the beam proton \cite{MS}.
Furthermore, since
the Coulomb interaction between the protons and electrons
is spin-independent it cannot provide spin-flip transitions and,
 consequently, does not contribute to the polarization buildup.
The same argument, obviously, is valid in case of
antiproton scattering off a hydrogen target. Therefore, the 
authors of Ref.~\cite{MS} concluded that only the hadronic interaction 
can be used to produce polarized antiprotons on the basis of the
spin-filtering mechanism.

In contrast to the $NN$ case, the spin-dependent part of the $\bar pN$ 
interaction is poorly known experimentally at present. 
Therefore, to investigate the 
polarization buildup mechanism in $\bar p~^1$H scattering 
a new experiment is planned 
at CERN \cite{AD}. The stored antiprotons will be scattered off
a polarized $^1$H target in that experiment \cite{AD} and the polarization
of the antiproton beam will be measured at intermediate energies.
Some theoretical estimations \cite{DmitrievMS}
of the expected polarization effects were already performed
employing a specific model of the $\bar{p}p$ interaction.

In this context, it is interesting and useful to explore other hadronic
reactions as possible source for the antiproton polarization buildup too.   
Therefore, in the present work we study polarization effects in 
antiproton-deuteron (${\bar p} d$) scattering for beam energies up to
300 MeV. Besides the issue of polarization buildup for 
antiprotons, $\bar{p}$ scattering on a polarized deuteron,
if it will be studied experimentally, can be also used
as a test for our present knowledge of the ${\bar p}n$ and ${\bar p}p$ 
interactions.  
Our investigation is based on the Glauber-Sitenko theory for 
${\bar p}d$ scattering and it utilizes the ${\bar N}N$ interaction models 
developed by the J\"ulich group \cite{Hippchen,Hippchen1,Mull} 
as input for the elementary amplitudes. Since there are data on 
(unpolarized) total and differential $\bar p d$ cross sections in
the considered energy range we can examine the reliability of the 
Glauber approach via a direct comparison of our results to 
experimental information. 
In addition we also present results for polarization effects for the
$\bar NN$ system itself. With the $\bar NN$ potentials developed by 
the J\"ulich group we have conceptually rather different models 
at our disposal than the one used in Ref.~\cite{DmitrievMS} and
it will be instructive to see and compare the corresponding predictions.
Moreover, we consider here the ${\bar p}p$ as well as the ${\bar p}n$
case. 

Let us mention here for completeness that very recently new QED
calculations for polarization transfer in proton-electron 
(and antiproton-positron) elastic scattering
were performed \cite{Arenhoevel}. These calculations
predict very large polarization transfer from polarized 
electrons (positrons) to unpolarized protons (antiprotons) in elastic
proton-electron (antiproton-positron) scattering
at low beam energies, less than 20 keV \cite{Arenhoevel}.
On this basis new sources for polarized antiprotons are under discussion
\cite{Walcher}. On the other hand, 
according to a calculation presented in Ref.~\cite{MSS}, the effect of
polarization buildup is negligible in this reaction, even at small
relative velocites. A recent measurement performed at COSY \cite{Frankpc} 
intends to explore this method.
Finally, another method for buildup of  
polarized antiprotons at high energies, based on production of
${\bar \Lambda}(1115)$ 
and its subsequentional decay  ${\bar \Lambda}\to \pi^+ +\bar p$,
has been proposed in Ref.~\cite{Nurushev}. 

The paper is structured in the following way. 
In the next section we consider the spin structure of the total ${\bar p}d$
cross section using the optical theorem. 
In Sect. III expressions for the forward 
${\bar p}d$ elastic scattering amplitude are derived
in the impulse approximation and the formalism for the polarized total
$\bar p d$ cross sections is developed. 
The formalism for calculating the Coulomb-nuclear interference cross 
sections is presented in Sect. IV. The technical details 
for evaluating the Coulomb-hadronic interference cross sections for 
$\bar p d$ elastic scattering are summarized in Appendix A.
The $\bar p N$ interaction model of the J\"ulich group is briefly reviewed
in Sect. V. The amplitudes of this model are used as input for
our ${\bar p}d$ calculations. We also provide and discuss results for 
spin-dependent cross sections in the $\bar p p$ as well as $\bar p n$ channels. 
Numerical results for the ${\bar p}d$ reaction are presented and discussed
in section VI.
A short summary is provided in the last section.
        
\section {Phenomenology of the spin dependence of
 the total $\bar p d$ cross section}

Let us first consider only the purely hadronic part of the reaction 
amplitude. The modifications due to the presence of the Coulomb interaction
will be discussed in Sect. IV.
We use the optical theorem to derive the formalism for the total spin dependent 
$\bar p d$ cross sections. 
According to \cite{phillips} one has
\begin{equation}
\label{optth}
Im \frac {Tr(\hat \rho_i \hat F(0))}{Tr \hat\rho_i}=
\frac{k_{\bar p d}}{4\pi} \sigma_i , 
 \end{equation}
 where $\hat F(0)$ is the transition operator for $\bar p d$ elastic
 scattering at the angle $\theta=0$, $\rho_i$ is the initial
 spin-density matrix, $\sigma_i$ is the total cross section depending on
 the density matrix $\rho_i$, and $k_{\bar p d}$ is the momentum in the 
center-of-mass system (cms).

 The spin dependence  of the amplitude of the $\bar p d$ elastic scattering
 is the same as for the $pd$ elastic scattering. For collinear kinematics
 it contains four independent terms
 \cite{rekalo} and can be written as \cite{uzepan98} 
\begin{equation}
\label{fab}
\hat F_{\alpha\beta}(0)=g_1\delta_{\alpha\beta}+
(g_2-g_1)m_\alpha m_\beta+ig_3{\hat \sigma}_i
\epsilon_{\alpha\beta  i}+ i(g_4-g_3){\hat \sigma}_i m_i m_j
\epsilon_{\alpha\beta  j},
\end{equation}
where $\hat \sigma_i$ ($i=x,y,z$) are the Pauli spin matrices,
$\epsilon_{\alpha \beta \gamma}$ is the fully antisymmetric tensor,
$m_\alpha$ are the Cartesian components of a unit vector ${\bf m}$
pointing along the beam momentum, and $g_i$ $(i=1,\dots,4)$ are complex numbers
determined by the dynamics of the reaction. Let us put the $z$ axis along 
the vector ${\bf m}$, so that ${\bf m}=(m_x,m_y,m_z)=(0,0,1)$.
The $\bar p d$ elastic scattering amplitude is obtained by sandwiching 
the operator $\hat F_{\alpha\beta}$ between antiproton and deuteron spin states, 
\begin{eqnarray}
\label{tfi}
F_{\mu \lambda}^{\mu^\prime \lambda^\prime}\equiv <\mu^\prime\lambda^\prime|{\hat F}|\mu\lambda>=
\phi^+_{\mu^\prime}e_{\beta}^{(\lambda^\prime)^*}{\hat F}_{\alpha \beta}
e_\alpha^{(\lambda)}\phi_\mu,
\end{eqnarray}
 where $\phi_\mu$ ($\phi_{\mu^\prime }$) is the Pauli spinor for the initial (final)
 antiproton with the spin projection $\mu$ ($\mu\prime$) and
 $e_\alpha^{(\lambda)}$ ($e_\beta^{(\lambda^\prime)}$) is the polarization vector of
 the initial (final) deuteron with the spin projection $\lambda$ ($\lambda^\prime)$.
 
 When choosing different initial polarization states in Eq.~(\ref{optth}),
 described by the product of the spin density matrices of
 the antiproton $\rho^{\bar p}$ and
 the deuteron ${\hat \rho^d}$, $\rho=\rho^{\bar p}\rho^d$,
 one can derive from Eqs.~(\ref{optth}) and (\ref{fab}) different total
 spin-dependent
 cross sections in terms of the forward amplitudes $g_i$.
 For example, for unpolarized antiprotons and unpolarized deuterons one has 
$\rho=\frac{1}{2}\times\frac{1}{3}$, and one finds from Eq.~(\ref{optth}) 
 the unpolarized total cross section to be 
 \begin{equation}
\label{sigma0}
 \sigma_0=\frac{2 \pi}{k_{\bar p d}}Im(g_1+g_2) \ .
\end{equation}
 In general the spin density matrices are
\begin{equation}
\label{rhop}
\rho^{\bar p}=\frac{1}{2}(1+{\bf  P}^{\bar p}{\hat {\bfg \sigma}})
\end{equation}
 for the antiproton and
\begin{equation}
\label{rhod}
\rho^{d}=\frac{1}{3}+ \frac{1}{2}S_jP_j^d+\frac{1}{9} S_{jk}P_{jk}^d
\end{equation}
 for the deuteron. Here $S_j$ is the spin-1 operator, 
 ${ P}_j^d$ and $P_{jk}^d$ $(j,k=x,y,z)$ are the vector
 and tensor polarizations of the deuteron, and
 $S_{jk}=(S_jS_k+S_kS_j-\frac{4}{3}\delta_{jk})$ is the spin-tensor operator.

Using Eqs.~(\ref{fab}),
 (\ref{rhop}) and (\ref{rhod}) one can find from Eq.~(\ref{optth}),
\begin{eqnarray}
\label{trace}
Tr \{\rho^{\bar p}\times\rho^d {\hat F}(0)\}=
 \frac{1}{2}(g_1+g_2)+P_x^{\bar p}P_x^dg_3+ P_y^{\bar p}P_y^dg_3+
P_z^{\bar p}P_z^dg_4+\nonumber \\
+\frac{1}{9}P_{xx}^d(g_1-g_2) +\frac{1}{9}P_{yy}^d(g_1-g_2) +
\frac{5}{18}P_{zz}^d(g_1-g_2).
\end{eqnarray}

 As seen from this formula, 
 if the initial antiproton is polarized with the polarization vector
 ${\bf  P}^{\bar p}$ and the deuteron has the 
 polarization vector ${\bf P}^d$, then non-zero terms $\sigma_i$ arise in the
 right-hand side of Eq.~(\ref{optth}) for parallel (or antiparallel) orientation
 of the vectors ${\bf P}^{\bar p}$ and ${\bf P}^d$.
  For the  tensor
 polarization of the deuteron only the diagonal components of the
 polarization tensor $P_{xx}^d$, $P_{yy}^d$ and $P_{zz}^d$ are connected with
 non-zero total cross section.
 The cross sections associated with  $P_{xx}^d$ and $P_{yy}^d$
 are the same and determined  by $\frac{1}{9}(g_1-g_2)$,
 whereas the cross section connected with the $P_{zz}^d$
 component is given by $\frac{5}{18}(g_1-g_2)$.
 Taking into account the relation $P_{xx}^d+P_{yy}^d+P_{zz}^d=0$,
 one can find from Eq.~(\ref{trace}) that
 the total tensor cross section can be connected only with the $P_{zz}^d$
 component.

  All other combinations of polarizations of the antiproton
 and/or deuteron, namely, polarized antiproton,
 vector-polarized deuteron, and polarized antiproton -- tensor-polarized 
 deuteron, give zero contribution to the total 
 cross section due to parity conservation.   

 Summarizing the above results, the total polarized ${\bar p} d$ section
 can be written as
\begin{equation}
\label{totalspin}
\sigma=\sigma_0+\sigma_1{\bf P}^{\bar p}\cdot {\bf P}^d+
 \sigma_2 ({\bf P}^{\bar p}\cdot {\bf m}) ({\bf P}^d\cdot {\bf m})+
\sigma_3 P_{zz}^d, 
\end{equation}
where $\sigma_0$ is given by Eq.~(\ref{sigma0}), and the other coefficients
are:
\begin{eqnarray}
\nonumber          
\sigma_1&=&\frac{4\pi}{k_{\bar p d}}Im g_3,\\
\nonumber 
\sigma_2&=&\frac{4\pi}{k_{\bar p d}}Im (g_4-g_3),\\
\label{sigma3}
\sigma_3&=&\frac{4\pi}{k_{\bar p d}}Im \frac{(g_1-g_2)}{6}.
\end{eqnarray}
 One can find from Eq.~(\ref{totalspin}) that only the cross sections 
$\sigma_1$ and $\sigma_2$
 are connected with the spin-filtering mechanism and,
 therefore, determine the rate of the
 polarization buildup in the scattering of unpolarized antiprotons off polarized
 deuterons. 
(More precisely,
 the rate of polarization buildup is determined by the ratio 
 $\sigma_1/\sigma_0$
 for transversal polarization and by $(\sigma_2-\sigma_1)/\sigma_0$ for
 longitudinal polarization of target and beam in the storage ring.)
 The tensor cross section $\sigma_3$ is not connected with the
 polarization of the beam and, therefore, is not relevant 
 for the spin-filtering. However, this cross section, as well as the 
 unpolarized cross section $\sigma_0$,
 determines the lifetime of the beam. When changing the sign of the tensor
 polarization $P_{zz}^d$, one may change the beam lifetime. 

\section{$\bar p d$ elastic scattering at forward angles}

\subsection {Glauber theory}
Within the Glauber theory the amplitudes
for the elastic ($\bar p d\to \bar p d$) and breakup 
($\bar p d\to \bar p np $)
reactions are given by the following matrix element  
\begin{equation}
\label{glafi}
F_{if}({\bf q})= <f|F({\bf q}, {\bf s})|i>,
\end{equation}
calculated between definite initial $|i>$ and final $|f>$ states of
the two-nucleon system. Here the transition operator is
\begin{equation}
\label{Ffi}
F({\bf q}, {\bf s})=\exp{(\frac{1}{2}i{\bf q}\cdot{\bf s})}f_{\bar p p}({\bf q})+
\exp{(-\frac{1}{2}i{\bf q}\cdot{\bf s})}f_{\bar p n}({\bf q})+
\frac{i}{2\pi k_{\bar p d}}\int \exp{(i{\bf q}'\cdot{\bf s})}
f_{\bar p n}({\bf q'}+\frac{1}{2}{\bf q})
f_{\bar p p}(-{\bf q'}+\frac{1}{2}{\bf q})d^2{\bf q}'.
\end{equation}
In Eqs.~(\ref{glafi}) and (\ref{Ffi}) ${\bf q}$ is the transferred momentum,
${\bf s}$ is the impact parameter, and $f_{\bar p N}({\bf q})$ ($N=p, n$) is the 
$\bar p N$ scattering amplitude.
The amplitude of elastic $\bar p d$ scattering can be expressed via 
the elastic form
factor of the deuteron, $S({\bf q})$, and the elementary amplitudes of
$\bar pN$ scattering. The differential scattering cross section for
elastic ($\bar p d \to \bar p d$) plus inelastic ($\bar p d \to \bar p pn$) 
scattering is calculated
within the closure approximation \cite{FrancoGlauber}. That allows one
to express the scattering cross section
 via $f_{\bar p N}({\bf q})$, 
$S({\bf q})$ and the deuteron wave function. 
Since the D-wave component
of the deuteron wave function becomes important only in the region of the 
first diffraction minimum of the differential $\bar p d$ cross section
\cite{mahalabi}, we neglect its contribution in the present calculations 
and take into account only the S-wave component.
  
 As seen from Eq.~(\ref{Ffi}), in the Glauber theory
 of multiple scattering of hadrons off the deuteron only
 single-scattering (first two terms on the right-hand side) and double-scattering 
 (third term on the right-hand side) mechanisms contribute to the transition
 amplitude. 
 In forward direction the single-scattering mechanism dominates.
 The corrections related to double-scattering effects produce the so-called 
 shadowing effect. As a result, the total unpolarized antiproton-deuteron 
 cross section $\sigma^{\bar p d}$ is not equal to the sum of the total 
 $\bar p p$ and $\bar p n$ cross sections, $\sigma^{\bar p p}$ and $\sigma^{\bar p n}$, 
 but is given by 
 \begin{equation}
\label{shadowing}
\sigma^{\bar p d}=\sigma^{\bar p p}+\sigma^{\bar p n}-\delta\sigma^d ,
\end{equation}
where $\delta\sigma^d$ stands for the corrections due to 
double-scattering effects.

\subsection{Impulse approximation}
    
In the impulse approximation (IA) (or single-scattering approximation) 
one can present the forward $\bar p d$ elastic scattering amplitude in the 
following form
\begin{equation}
\label{pdamp0}
F_{\mu \lambda}^{\mu' \lambda'}=
\frac{m_d}{m_N}\sqrt{\frac{s_{\bar p N}}{s_{\bar p d}}}
\sum_{\sigma \,\sigma'\, \sigma_N}
<\sigma'\mu'|f_{\bar p N}|\sigma\mu>
S_{\lambda\, \lambda'}^{\sigma \sigma'\sigma_N}({\bf Q}=0),
\end{equation}
 where $<\sigma'\mu'|f_{\bar p N}|\sigma\mu>$ is the $\bar p N$ scattering 
amplitude at zero degree, defined as in Ref.~\cite{bystricky}, 
 $S_{\lambda\, \lambda'}^{\sigma \sigma'\sigma_N}({\bf Q}=0)$ is the elastic
 form factor of the deuteron at zero transferred momentum ${\bf Q}$, 
 $s_{\bar p N}$ ($s_{\bar p d}$) is the invariant mass of the $\bar p N$
 ($\bar p d$) system, and 
$m_d$ ($m_N$) is the mass of the deuteron (nucleon).
 The sum in Eq.~(\ref{pdamp0}) runs over the 
 $z$ projections of the spin of the nucleon spectator $(\sigma_N)$, 
and of the initial ($\sigma$) and recoil ($\sigma'$) nucleons inside the deuteron.
The transition operator for the $\bar p N$ forward scattering
amplitude has the form \cite{bystricky}
\begin{equation}
\label{pNamp}
f_{\bar p N}=A_N+B_N ({\bfg \sigma}_1\cdot {\bfg \sigma}_2) +D_N
({\bfg \sigma}_1\cdot {\bf m})({\bfg \sigma}_2\cdot {\bf m}),
\end{equation}
where the matrix ${\bfg \sigma}_1$  (${\bfg \sigma}_2$)
acts on the spin state of the antiproton (nucleon) and $A_N$, $B_N$ and $D_N$ ($N=p,n$) 
are complex amplitudes \cite{bystricky}.

The deuteron elastic form factor at ${\bf Q}=0$ can be written as
\begin{equation}
 \label{dff}
 S_{\lambda\, \lambda'}^{\sigma \sigma'\sigma_N}({\bf Q}=0)=
\sum_{l\, m\, M_S\,M_S'}(\frac{1}{2}\sigma \frac{1}{2}\sigma_N|1M_S)
 (1M_S\,l\,m|1\lambda)
(\frac{1}{2}\sigma \frac{1}{2} \sigma_N|1M_S)
 (1M_S\,l\,m|1\lambda)(1M_S'\,l\,m|1\lambda') \, P_l,
\end{equation}
where $P_l$ is the relative weight of the S-wave ($l=0$) and D-wave $(l=2)$ 
components of the deuteron wave function with the normalization $P_0+P_2=1$.

Using Eq.~(\ref{fab}) one can find the invariant amplitudes $g_1,\dots,g_4$
 from the following transition  matrix elements  
\hbox{$<\mu'\lambda'|{\hat F}|\mu \lambda>$} of $\bar p d$ forward
 scattering \cite{rekalo}:
\begin{eqnarray}
\label{amprelations1}
<+\frac{1}{2}, \phantom{+}0|\hat F|+\frac{1}{2}, \phantom{+}0>&=& g_2 \nonumber \\
<+\frac{1}{2}, +1|\hat F|+\frac{1}{2}, +1>&=& g_1-g_4, \nonumber \\
 <+\frac{1}{2}, -1|\hat F|+\frac{1}{2}, -1>&=& g_1+g_4, \nonumber \\
<+\frac{1}{2}, -1|\hat F|-\frac{1}{2}, \phantom{+}0>&=& -\sqrt{2}g_3.
\end{eqnarray}
On the other hand, using Eqs.~(\ref{pdamp0}), (\ref{pNamp}) and (\ref{dff}),
one can express the transition matrix elements in terms of the $\bar pN$
scattering amplitudes as
\begin{eqnarray}
\nonumber
<+\frac{1}{2}, \phantom{+}0|\hat F|+\frac{1}{2}, \phantom{+}0>&=&A\,w,\\
\nonumber
<+\frac{1}{2}, +1|\hat F|+\frac{1}{2}, +1>&=&(A+B+D)\,w,\\
\nonumber
<+\frac{1}{2}, -1|\hat F|+\frac{1}{2}, -1>&=& (A-B-D)\,w,\\
\label{amprelations2}
<+\frac{1}{2}, -1|\hat F|-\frac{1}{2}, \phantom{+}0>&=&\sqrt{2}B\,w,
\end{eqnarray}
where $A=A_p+A_n$, $B=B_p+B_n$, $D=D_p+D_n$ and 
\begin{eqnarray}
\label{pspd}
w=\frac{m_d}{m_N}
\sqrt{\frac{s_{\bar pN}}{s_{\bar p d}}}(P_0-\frac{1}{2}P_2). 
\end{eqnarray}
From a comparison of Eqs.~(\ref{amprelations1}) with 
Eqs.~(\ref{amprelations2}) one can find that in the single-scattering 
approximation the invariant amplitudes can be written as
\begin{eqnarray}
\label{g4toab}
g_1=g_2= Aw&=&\phantom{-}\frac{1}{2}\left [M_1(0)+M_3(0)\right ]w,\nonumber \\
g_3=-Bw&=&-\frac{1}{2} M_2(0)w, \nonumber\\
g_4=-(B+D)w&=&\phantom{-}\frac{1}{2}\left [M_1(0)-M_3(0)\right ]w . 
\end{eqnarray}

Here we used the standard relations between the amplitudes 
$A_N$, $B_N$, and $D_N$ and the helicity amplitudes $M_i^N$ 
of $\bar p N$ scattering \cite{bystricky} together with the notation $M_i=M_i^p+M_i^n$.
Utilizing Eqs.~(\ref{sigma0}), (\ref{sigma3}) and (\ref{g4toab})  
the total cross sections are thus
\begin{eqnarray}
\nonumber 
\sigma_0^{IA}&=&\frac{2\pi}{k_{\bar p d}}Im\left [M_1^p(0)+M_3^p(0)+
M_1^n(0)+M_3^n(0)\right ] w = 
(\sigma_0^{\bar p p}+\sigma_0^{\bar p n}){\widetilde w},\\
\nonumber 
\sigma_1^{IA}&=&-\frac{2\pi}{k_{\bar p d}}Im \left [M_2^p(0)+M_2^n(0)\right ]w =
-(\sigma_1^{\bar p p}+\sigma_1^{\bar p n}){\widetilde w},\\
\nonumber 
\sigma_2^{IA}&=&\frac{2\pi}{k_{\bar p d}}Im\left [M_1^p(0)+M_2^p(0)-M_3^p(0)
+M_1^n(0)+M_2^n(0)-M_3^n(0) \right ] w =
-(\sigma_2^{\bar p p}+\sigma_2^{\bar p n}){\widetilde w},\\
\label{sigma3IA}
\sigma_3^{IA}&=&0,
\end{eqnarray}
where we used the relations \cite{bystricky}
\begin{eqnarray}
\nonumber
\sigma_0^{\bar p N}&=&\frac{2\pi}{k_{\bar p N}}Im\left [M_1^N(0)+M_3^N(0)\right ], \\
\nonumber 
\sigma_1^{\bar p N}&=&\frac{2\pi}{k_{\bar p N}}Im\left [M_2^N(0)\right ], \\
 \label{s2IA}
\sigma_2^{\bar p N}&=&-\frac{2\pi}{k_{\bar p N}}Im\left [M_1^N(0)+M_2^N(0)-M_3^N(0)\right ],
\end{eqnarray}
and the fact that the cms momentum in the $\bar pN$ system,
$k_{\bar p N}$,
is related to the ${\bar p d}$ cms momentum, $k_{\bar p d}$, by
\begin{equation}
\frac{k_{\bar pd}}{k_{\bar p N}}=\frac{m_d}{m_N}
\sqrt{\frac{s_{\bar pN}}{s_{\bar p d}}} \ ,
\end{equation}
for equal ($\bar p$) beam energies $T_{lab}$ in the reactions $\bar p N$ 
and $\bar p d$. The quantity $\widetilde w$ in Eqs.~(\ref{sigma3IA}) is defined by 
$\widetilde w=P_0-\frac{1}{2}P_2$.

One can see from Eqs.~(\ref{sigma3IA}) that in the impulse 
approximation all total cross sections are additive, i.e. given by the sum of the
corresponding cross sections on the proton and neutron.
While this result is obvious for the total unpolarized ${\bar p d}$
cross section $\sigma_0$, it was not expected for the
spin-dependent cross sections, especially, in view of the
opposite sign of the $\bar p d$ cross sections $\sigma_1$ and $\sigma_2$
with regard to $\sigma_1^{\bar p N}$ and $\sigma_2^{\bar p N}$, respectively.
 
\subsection{Shadowing effects}

As already mentioned above, the double-scattering mechanism dominates at large 
scattering angles, while its relative contribution decreases when approaching 
the scattering angle $\theta = 0$. 
The numerical calculations of the forward amplitude of $\bar p d$ elastic 
scattering, which will be presented below, demonstrate that the inclusion of 
the double-scattering mechanism reduces the total unpolarized cross section 
by around 15~\% at energies 50-300 MeV as compared to the result obtained 
within the single-scattering approximation 
If one adopts the approximaton given in Ref.~\cite{Alberi} then 
the effects of double-scattering for polarized $\bar p d$ cross sections 
should be likewise around 15-20 \% in the energy region in question. 
Thus, we expect them to be significantly smaller than the variations in
the predictions due to the uncertainties in the spin-dependence of the 
elementary $\bar p N$ interaction. Therefore, we do not consider the 
double-scattering effects on the spin-dependent cross sections in the 
present investigation, which anyway has an exploratory character. 
Note that 
an accurate evaluation of the contribution of double-scattering effects to
polarized cross sections requires to consider the D-wave component of the 
deuteron as well as the angular dependence of all (ten) $\bar pN$ helicity 
amplitudes \cite{Alberi1} and is therefore rather tedious. 

\section{Coulomb-nuclear interference}

The total polarized cross sections including the Coulomb
interaction can be written as the sum of the purely hadronic 
contributions $\sigma_i^h$, 
the Coulomb-nuclear interference terms $\sigma^{int}_i$ and the 
pure Coulomb contribution $\sigma^C_i$: 
\begin{eqnarray}
\label{sigma0tot}
 \sigma_0&=&\sigma_0^h+\sigma_0^{int}+\sigma_0^C,\nonumber \\
\label{sigma1tot}
\sigma_1&=&\sigma_1^h+\sigma_1^{int}+\sigma_1^C,\nonumber \\
\label{sigma2tot}
\sigma_2&=&\sigma_2^h+\sigma_2^{int}+\sigma_2^C,\nonumber \\
 \label{sigma3tot}
\sigma_3&=&\sigma_3^h+\sigma_3^{int}+\sigma_3^C,
 \end{eqnarray}
where $\sigma_3$ is only present in case of the $\bar p d$ reaction.
The hadronic contributions are those discussed in detail in the 
two preceeding sections. Note, however, that from now onwards we label 
the corresponding quantities with the superscript ``h'' for the sake 
of clarity.  

As was found in Refs.~\cite{MS,HOM}, the interference between
the Coulomb amplitude and the hadronic amplitudes in the total 
spin-dependent cross section of $pp$ scattering plays an important role 
in the spin-filtering mechanism. When taken
into account together with the purely hadronic total cross section
this interference improves significantly the agreement between
the theory of spin-filtering and the data of the FILTEX experiment 
\cite{MS,NNNP,HOM}.  

Due to the singularity at $\theta\to 0$, Coulomb effects in the total 
cross section cannot be taken into account by means of the optical 
theorem. Therefore, in order to obtain the Coulomb-nuclear
 interference cross section for elastic scattering
 one has to perform an integration over the scattering
 angle for terms like $Re f^C(\theta){f^h}^*(\theta)$
 or $Im f^C(\theta){f^h}^*(\theta)$, where $f^C$ is the
Coulomb scattering amplitude, and $f^h$ is the amplitude
of the purely hadronic interaction 
 modified by the Coulomb interaction \cite{Landau}.
 As was shown in \cite{MS} and \cite{NNNP}, in such an
 integration the lower limit for the scattering angle $\theta$
 has to be taken as $\theta_{acc}$ (with $\theta_{acc}<<1$),
 because scattering events at lower angles, where the
 projectiles stay in the beam, do not lead to a beam polarization.
 
In the Glauber theory Coulomb effects can be taken into account by the method
of Ref.~\cite{Lesniak} in which the elementary eikonal $pp$ phase
is taken as sum of the purely strong and purely Coulomb phases. 
It is rather obvious that Coulomb effects appear in the
total $\bar p d$ cross section only due to the presence of the pure Coulomb 
term in the $\bar p p$ elastic scattering amplitude (see Eq.~(\ref{purekul})).
However, there is also an interference effect between the Coulomb amplitude 
and the $\bar pn$ scattering amplitude, as can be seen immediately from the 
expressions for the single-scattering approximation given below.  

\subsection{ $\bar p p$ scattering}

When calculating the Coulomb total cross section
and the Coulomb-nuclear interference cross sections for the $\bar p p$
system we follow Ref.~\cite{MS}, where these effects were
considered for $pp$ scattering. Here we take into account the 
difference in the electric charge between antiproton and proton, and we drop 
the exchange term $f^C(\pi-\theta)$, specific for $pp$ scattering.
The Coulomb scattering amplitude for $\bar p p$ is
\begin{equation}
 \label{purekul}
f^C(\theta)=\frac{\alpha}{4vk_{\bar p p}\sin^2{\theta/2}}\exp{\left\{ i\frac{\alpha}{v}
\ln{\sin{\frac{\theta}{2}}+2i\chi_0} \right \}},
\end{equation}
where $\alpha$ is the fine
structure constant and $v$ ($k_{\bar p p}$) the velocity (momentum) of the antiproton in 
the $\bar p p$ cms. The Coulomb phase $\chi_0$ is
\begin{equation}
\label{chi0}
\chi_0=\arg \Gamma{\bigl (1-\frac{i\alpha}{2v}\bigr )}.
\end{equation}
 The cross sections were considered in Ref.~\cite{MS} under the assumption
 that the beam acceptance satisfies the following condition:
 $\theta_{acc}\ll \alpha/(vk_{\bar p p}{\overline{f_h}})$, where ${\overline{f_h}}$
 is the typical magnitude of the hadronic amplitude. Within this assumption
 the total Coulomb cross section was estimated in \cite{MS} to be 
\begin{equation}
\label{sigma0c}
 \sigma_0^C\approx \pi\alpha^2/(vk_{\bar p p}\theta_{acc})^2.
\end{equation}
In contrast to $pp$ scattering, for the $\bar p p$ interaction the spin-dependent 
Coulomb cross sections $\sigma_1^C$ and 
$\sigma_2^C$ are zero, because there is no 
antisymmetrization term in the elastic $\bar pp$ scattering amplitude.
 The interference terms $\sigma_1^{int}$ and $\sigma_2^{int}$ 
 obtained in \cite{MS} in the logarithmic approximation (see Eq.~(18) therein), 
 have a fairly smooth dependence on $\theta_{acc}$,
 namely of the form $\ln{\sin{\theta_{acc}}/2}$. 
 Adapting the formalism of \cite{MS} for the $\bar p p$ case we obtain
 the following expressions for the contribution of the Coulomb-nuclear 
 interference terms to the spin-dependent cross sections: 
\begin{eqnarray}
\nonumber
\sigma_0^{int} =-\frac{2\pi}{k_{\bar pp}}
\Bigl \{
 \cos{2\chi_0}\bigl [-\sin{\Psi} Re [ M_3^p(0)+M_1^p(0)]
 +(1-\cos{\Psi})Im [M_3^p(0)+M_1^p(0)]\bigr ] - \\
\nonumber 
-\sin{2\chi_0}\bigl [\sin{\Psi} Im [M_3^p(0)+M_1^p(0)]
 +(1-\cos{\Psi})Re [M_3^p(0)+M_1^p(0)]\bigr ] \Bigr \},\\
\nonumber
\sigma_1^{int} = -\frac{2\pi}{k_{\bar pp}}
\Bigl \{
\cos{2\chi_0} \bigl [ -\sin{\Psi} Re M_2^p(0)
 +(1-\cos{\Psi})Im M_2^p(0) \bigr ] - \\ 
\nonumber 
-\sin{2\chi_0}\bigl [ \sin{\Psi} Im M_2^p(0)
 +(1-\cos{\Psi})ReM_2^p(0)\bigr ]
\Bigr \}, \\
\label{CNpp2}
\sigma_2^{int}=-\frac{2\pi}{k_{\bar pp}}
\Bigl \{
\cos{2\chi_0}\bigl [-\sin{\Psi} Re [ M_3^p(0)-M_1^p(0)-M_2^p(0)]+
\nonumber \\
 +(1-\cos{\Psi})Im [M_3^p(0)-M_1^p(0)-M_2^p(0)] \bigr ] + \nonumber \\ 
+\sin{2\chi_0}\bigl [\sin{\Psi} Im [M_3^p(0)-M_1^p(0)-M_2^p(0)]
 -(1-\cos{\Psi})Re [M_3^p(0)-M_1^p(0)-M_2^p(0)]\bigr ]
\Bigr 
\}, 
\end{eqnarray}
where $\Psi=-\frac{\alpha}{v}\ln{\sin{{\theta_{acc}}/{2}}}$ and
$M_i^p(0)$ are the  
hadronic $\bar p p$ helicity amplitudes, modified by the Coulomb interaction.
One can see that in the limit $\chi_0\to 0$
 Eqs.~(\ref{CNpp2}) coincide
 with Eq.~(18) of Ref.~\cite{MS}. We should note that our definition for
 the cross section $\sigma_2$ differs from that in Ref.~\cite{MS}: our $\sigma_2$ 
 is equal to $\sigma_2-\sigma_1$ as definined in Eq.~(2) of Ref.~\cite{MS}.
It is worth to note that at sufficiently high beam energies (above $\approx$ 50
MeV) one has $\cos {2\chi_0}\approx 1$ and $\sin {2\chi_0}\approx 0$.
    As one can see from Eqs.~(\ref{CNpp2}), in this case the total $\bar pp$ 
    interference cross sections $\sigma_i^{int}$ are determined by $Re [M_1^p(0)+M_3^p(0)]$, 
    $Re M_2^p(0)$ and $Re [M_1^p(0)+M_2^p(0)- M_3^p(0)]$ for $i=0$, 1 and 2,
    respectively, whereas the purely hadronic total $\bar p p$ cross sections 
    $\sigma_i^h$ are given by the corresponding imaginary parts of those
    amplitude combinations, see Eqs.~(\ref{s2IA}). 

\subsection{$\bar p d$ elastic scattering }
\label{pditerf}
 
Also for $\bar p d$ the total Coulomb cross section contributes only for $i=0$. 
 In order to calculate the Coulomb-hadronic interference cross sections for
 $\bar p d$ one needs the elastic scattering amplitudes beyond the
 collinear kinematics, because it is necessary to perform an integration 
 over the scattering angle.
 Therefore, the full spin structure of the $\bar p d$ scattering amplitude which
 consists of twelve independent terms, has to be considered. Details of
 the formalism for the general case are summarized in Appendix A.
The final formulae for the polarized interference cross sections are
Eqs.~(\ref{intsecpd}).

In order to obtain $\sigma_i^{int}$ within the impulse approximation one has 
to insert Eqs.~(\ref{g4toab}) into Eqs.~(\ref{collinearF}) and after that
 use it in Eqs.~(\ref{intsecpd}). 
The Coulomb amplitude for $\bar p d$
scattering in the impulse approximation at $\theta \ll 1$ is
$ \left (F^C\right)_{\mu \lambda}^{\mu'\lambda'}\approx\delta_{\mu \mu'}\delta_{\lambda \lambda'}
\frac{k_{\bar p d}}{k_{\bar p p}}f^C(\theta){\widetilde w}$, where
$f^C(\theta)$ is defined in Eq.~(\ref{purekul}).
Thus, the Coulomb-nuclear interference contribution to the total
$\bar p d$ cross sections, $\sigma_i^{int}$,
($i=0,\dots, 2$) follows from Eqs.~(\ref{CNpp2}) with
the replacement $M_i^p(0)\to M_i^p(0)+M_i^n(0)$.
Furthermore, Eqs.~(\ref{CNpp2}) should 
be multiplied by the factor $(k_{\bar p d}/k_{\bar p p}){\widetilde w}$
and $\sigma_1^{int}$ and $\sigma_2^{int}$ will change their signs.
Note that the cross sections $\sigma_3^h$ and $\sigma_3^{int}$ are
equal to zero in the single-scattering approximation, because the hadronic 
amplitude $g_1-g_2$ vanishes in this approximation. 

\section{Results for the $\bar NN$ system}

In the present investigation we use two $\bar N N$ models
developed by the J\"ulich group. 
Specifically, we use the models A(BOX) introduced in Ref.~\cite{Hippchen}
and D described in Ref.~\cite{Mull}.
Starting point for both models is the full Bonn $NN$ potential~\cite{MHE};
it includes not only traditional one-boson-exchange diagrams but also
explicit $2\pi$- and $\pi\rho$-exchange processes as well as virtual
$\Delta$-excitations. The G-parity transform of this meson-exchange
$NN$ model provides the elastic part of the considered $N\bar N$ interaction 
models. 
In case of model A(BOX) \cite{Hippchen} (in the following 
referred to as model A) 
a phenomenological \hbox{spin-}, isospin- and energy-independent 
complex potential of Gaussian form is added to account for the 
$\bar N N$ annihilation. It contains only three free parameters (the range
and the strengths of the real and imaginary parts of the annihilation
potential), fixed in a fit to the available total and integrated
$\bar N N$ cross sections. 
In case of model D \cite{Mull}, the most complete $N\bar N$ model of the 
J\"ulich group, the $N\bar N$ annihilation into 2-meson decay 
channels is described microscopically, including all possible
combinations of $\pi$, $\rho$, $\omega$, $a_0$, $f_0$, $a_1$, $f_1$, 
$a_2$, $f_2$, $K$, $K^+$ -- see Ref. \cite{Mull} for details --
and only the decay into multi-meson channels is simulated by 
a phenomenological optical potential. 

\begin{figure}[t]
%
\includegraphics{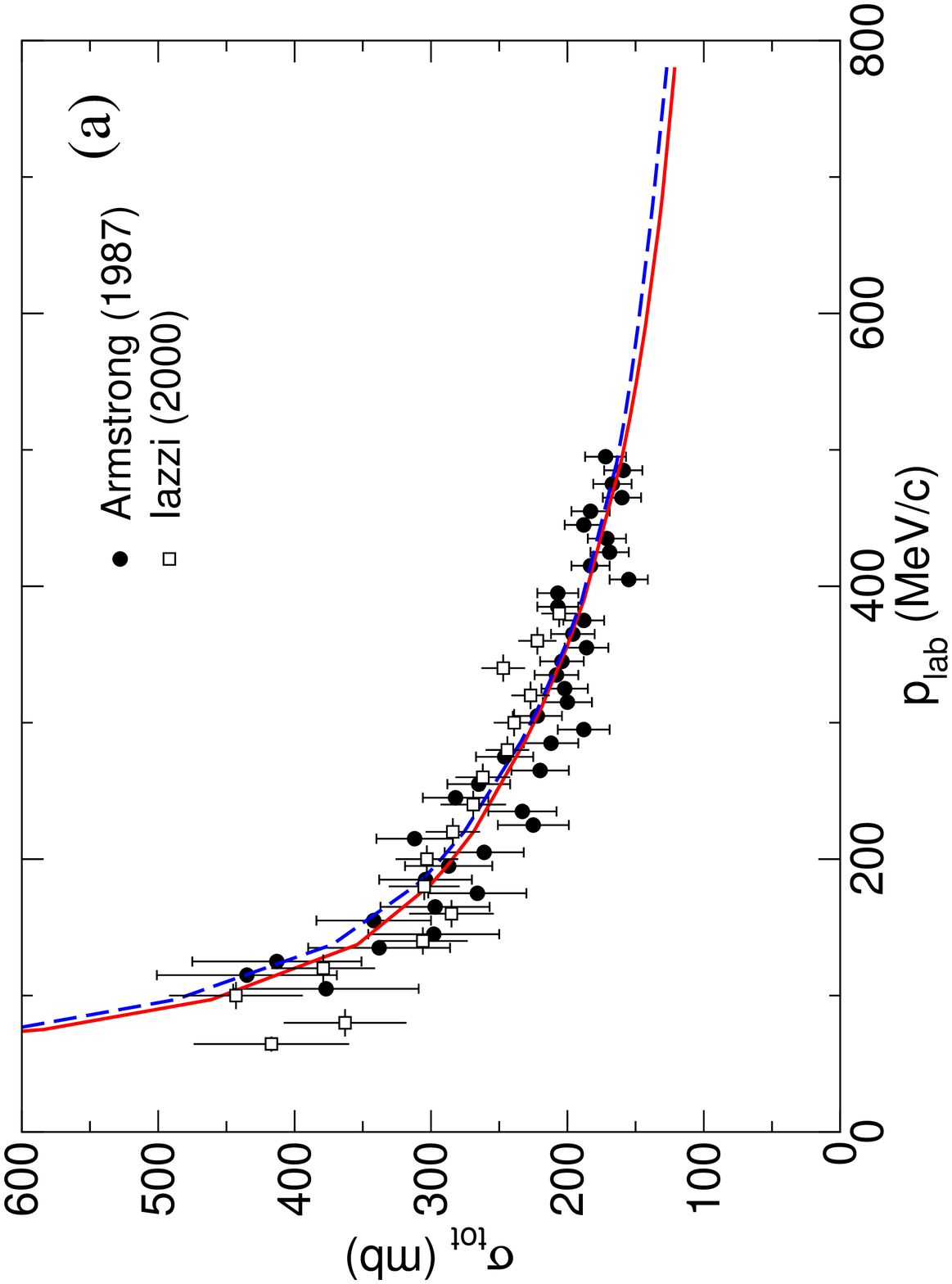}
\includegraphics{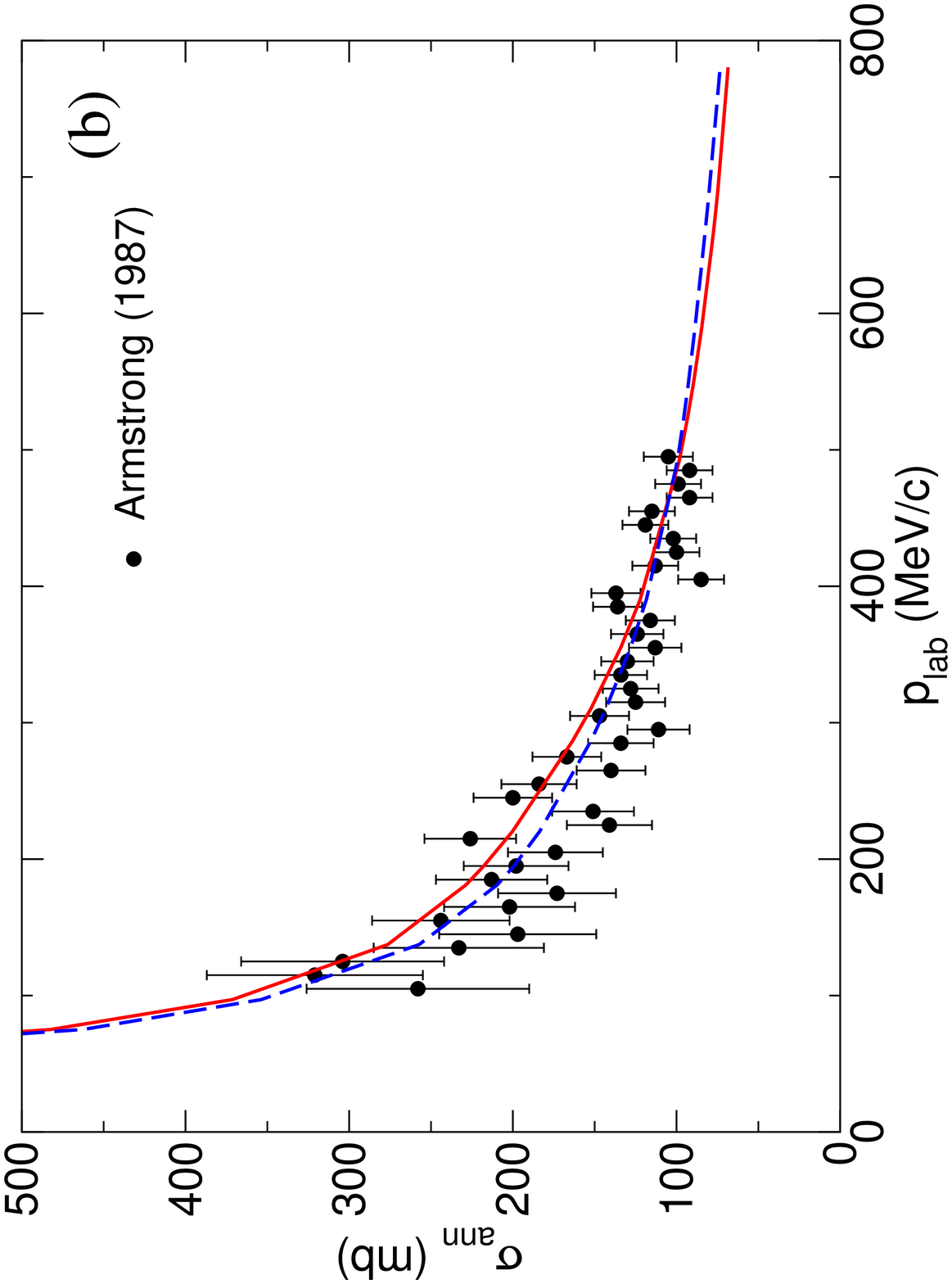}
\vskip 5.5cm
\caption{Total (a) and integrated annihilation (b) $\bar n p$ 
 cross sections versus beam momentum. The lines show predictions by the 
 J\"ulich $\bar NN$ models A (dashed line) and D (solid line).     
 Data are taken from Refs.~\cite{Arm87} (circles) and \cite{Iaz00}
 (squares).
}
\label{totpn}
\end{figure}
 
Results for the total and integrated elastic ($\bar p p$) and 
charge-exchange ($\bar p p \to \bar nn$) cross sections and also 
for angular dependent observables for both models can be found in 
Refs.~\cite{Hippchen,Mull}. Evidently, with model A as well as with
D a very good overall reproduction of the low- and intermediate 
energy $\bar N N$ data was achieved. 
Moreover, exclusive data on several $p\bar p$ 2-meson and even
3-meson decay channels are described with fair quality
\cite{Hippchen1,Mull,Betz}. 
Recently, it has been shown that the 
$N\bar N$ models of the J\"ulich group can also explain successfully
the near-threshold enhancement seen in the $\bar pp$ mass 
spectrum of the reactions $J/\Psi \to \gamma \bar pp$ \cite{Sibi1}, 
$J/\Psi \to \omega\bar pp$ \cite{Sibi3} and $B^+ \to K^+ \bar pp$ 
\cite{Sibi2} and in the $e^+e^-\to \bar pp$ 
cross section \cite{Sibi4}. 

As already mentioned in the Introduction, the spin dependence of the
$\bar NN$ interaction is not well known. There is a fair amount of
data on analyzing powers,
for $\bar pp$ elastic as well as for 
$\bar pp \to \bar nn$ charge-exchange scattering, cf. 
Ref.~\cite{Klempt} for a recent review. 
However, with regard to other spin-dependent observables
there is only scant information on the depolarization $D_{nn}$
and also on $K_{nn}$. Moreover, those data are of rather limited 
accuracy so that they do not really 
provide serious constraints on the $\bar NN$ interaction. 
The predictions
of the J\"ulich models A and D are in reasonable agreement with
the experimental polarizations up to beam momenta of 
$p_{lab}\approx 550$ MeV/c as can be seen in Ref.~\cite{Mull}.
In fact, model A gives a somewhat better account of the data and 
reproduces the measured $\bar pp$ polarizations even 
quantitatively up to $p_{lab}\approx 800$ MeV/c
($T_{lab} \approx 300$ MeV). 
We consider both models here because it allows us to 
illustrate the influence of uncertainties in the spin-dependence 
of the $\bar NN$ interaction on the spin-dependent cross
sections for the $\bar NN$ as well as the $\bar p d$ systems. 
In this context let us mention that a partial-wave analysis 
of $\bar pp$ scattering has been performed by the Nijmegen
Group \cite{Timmermans} which, in principle, would allow to pin 
down the spin-dependence of the $\bar NN$ interaction. However, 
the uniqueness of the achieved solution was disputed in Ref.~\cite{Richard}. 
Moreover, the actual amplitudes of the Nijmegen analysis
are not readily available and, therefore, cannot be 
utilized for the present investigation. 

For the computation of $\bar p d$ scattering we also need
the $\bar p n$ amplitude. For this system, a purely isospin $I=1$ 
state, there is no experimental information. 
But there are data for the $\bar n p$ channel \cite{Arm87,Iaz00}, 
which is identical to the former under the assumption of isospin symmetry.
A comparison of our model results with those data is
presented in Fig.~\ref{totpn}. Obviously the predictions of 
the J\"ulich models are in nice agreement with the experimental 
information on the $\bar n p$ interaction too, despite the fact 
that those total and annihilation cross sections have not been 
included in the fitting procedure. 


\begin{figure}
\includegraphics{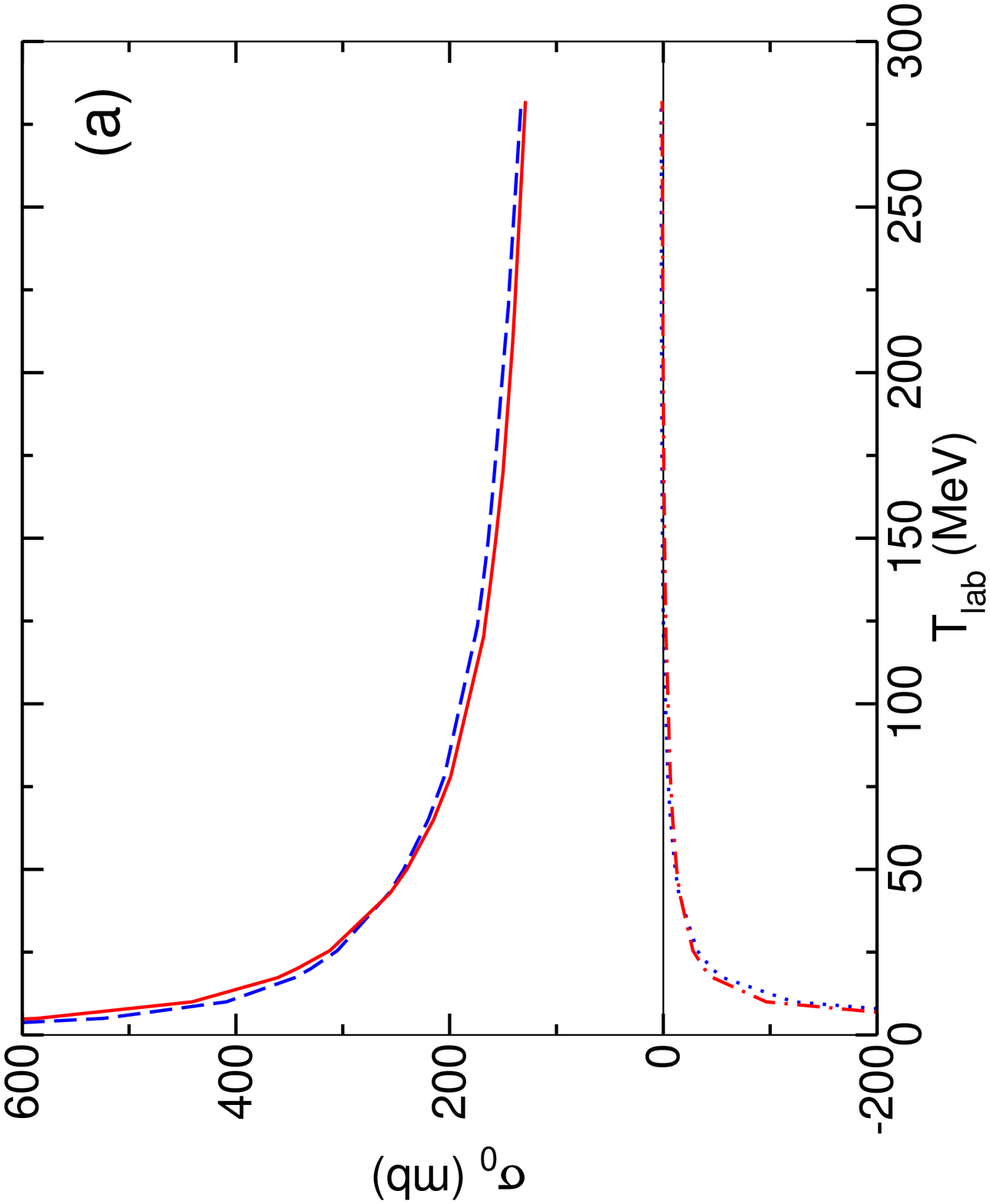}
\includegraphics{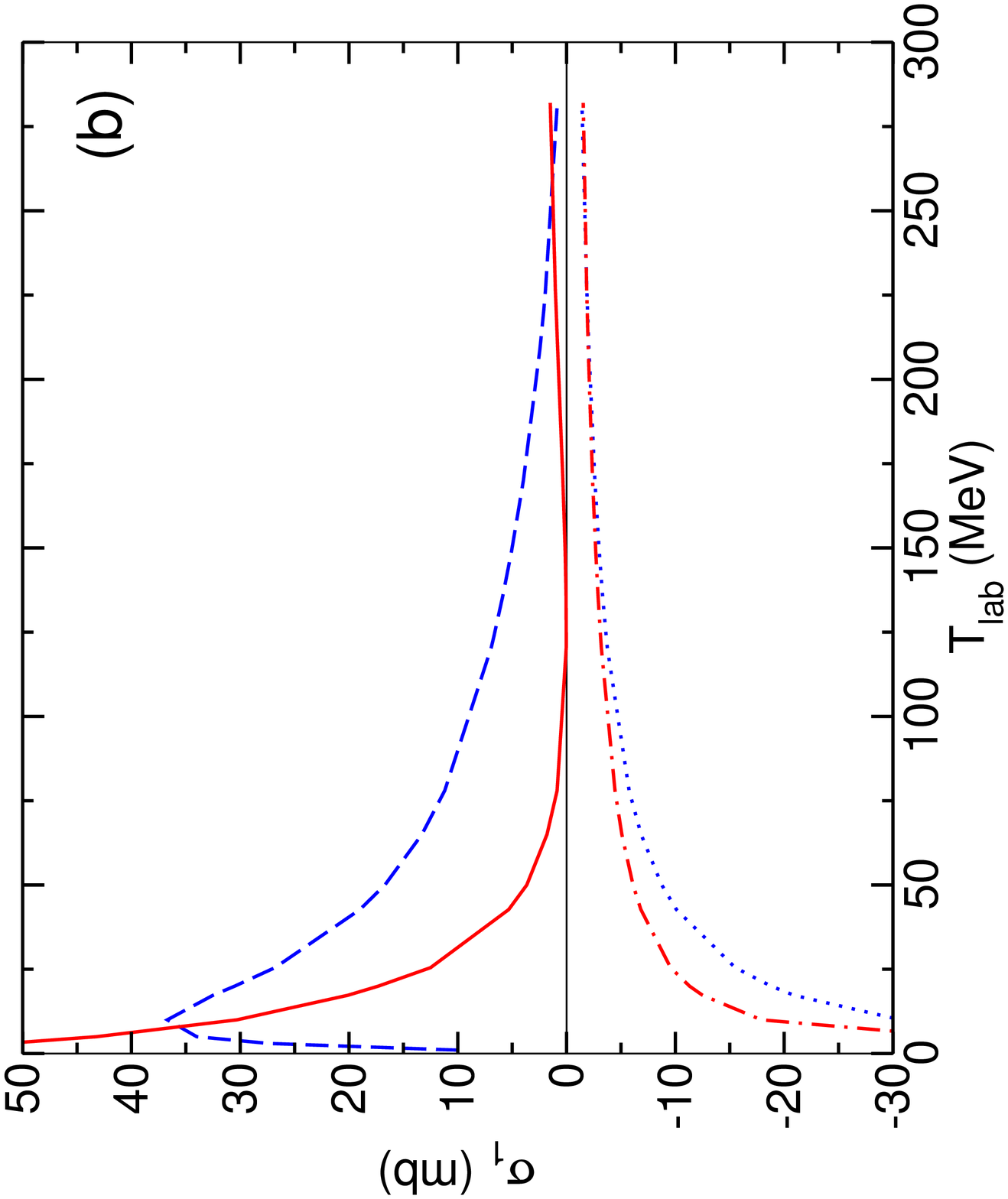}
\includegraphics{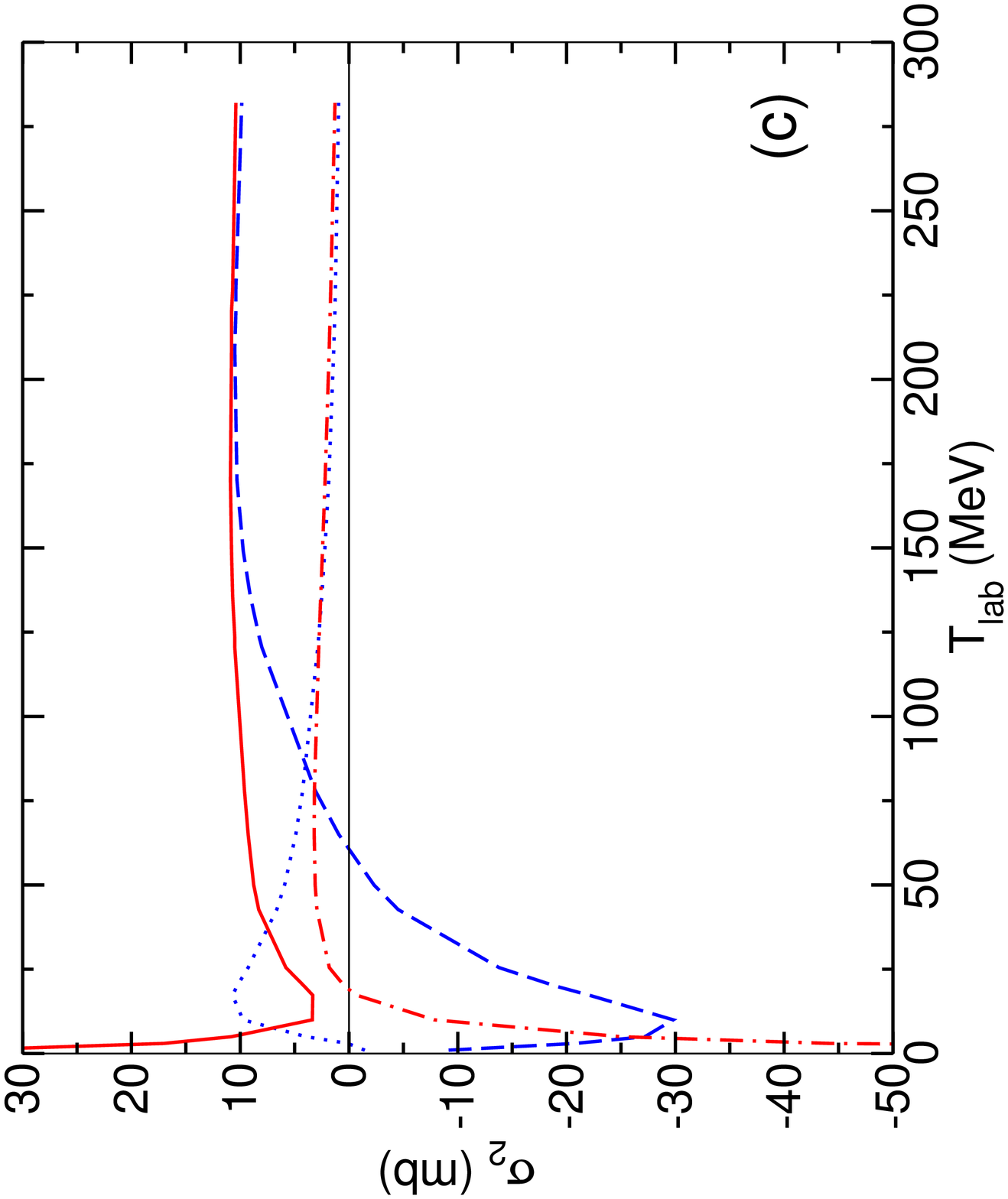}
\vskip 10.5cm
\caption{Total $\bar p p$ cross sections $\sigma_{0}$ (a), 
$\sigma_{1}$ (b), and $\sigma_{2}$ (c) 
versus the antiproton laboratory energy $T_{lab}$. 
Results based on the purely hadronic amplitude,  
$\sigma^h_i$, (model D: solid line, model A: dashed line) 
and for the Coulomb-nuclear interference term, 
$\sigma^{int}_i$, (D: dash-dotted line, A: dotted line) 
are shown.
}
\label{totppDA}
\end{figure}

\begin{figure}
\includegraphics{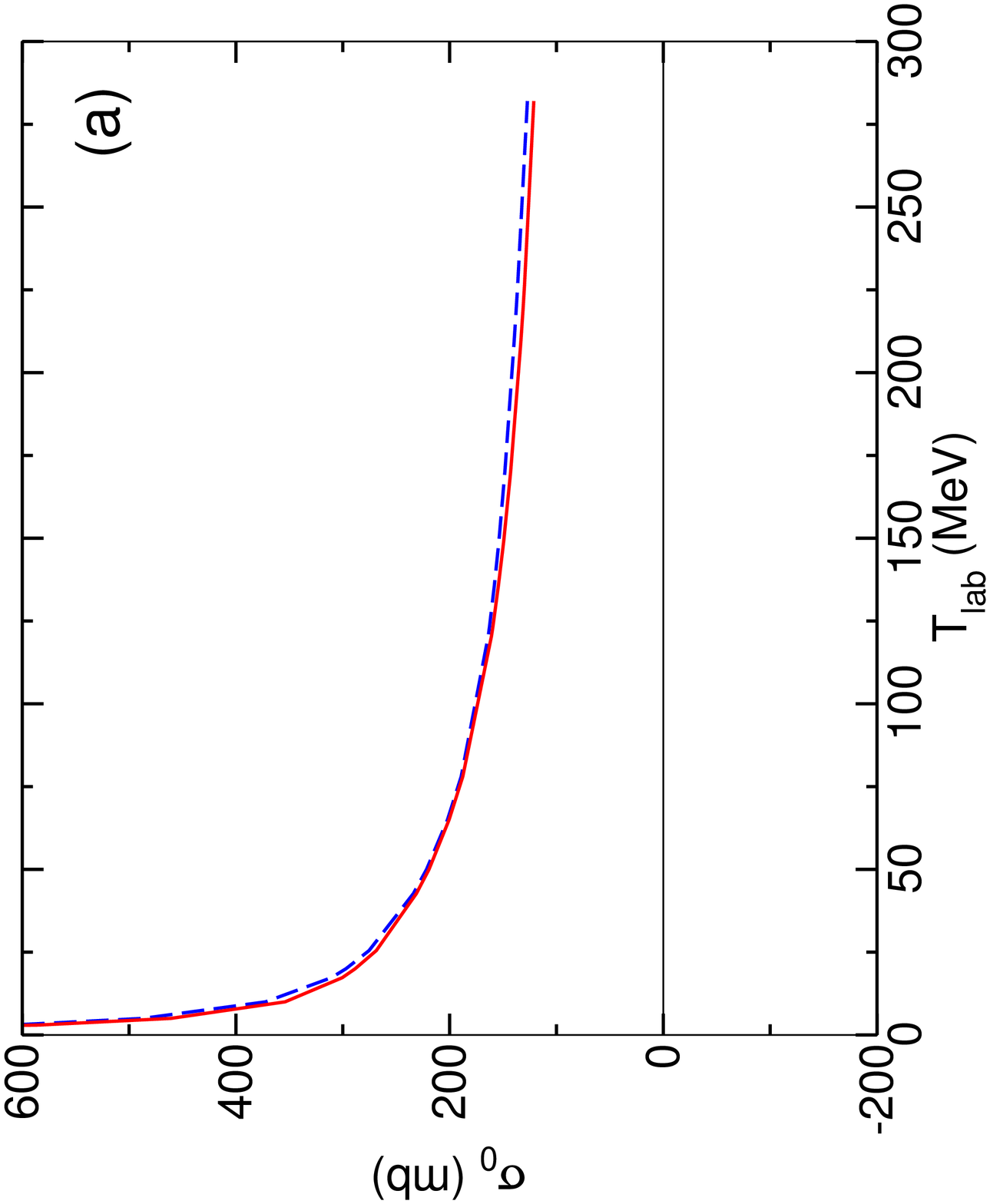}
\includegraphics{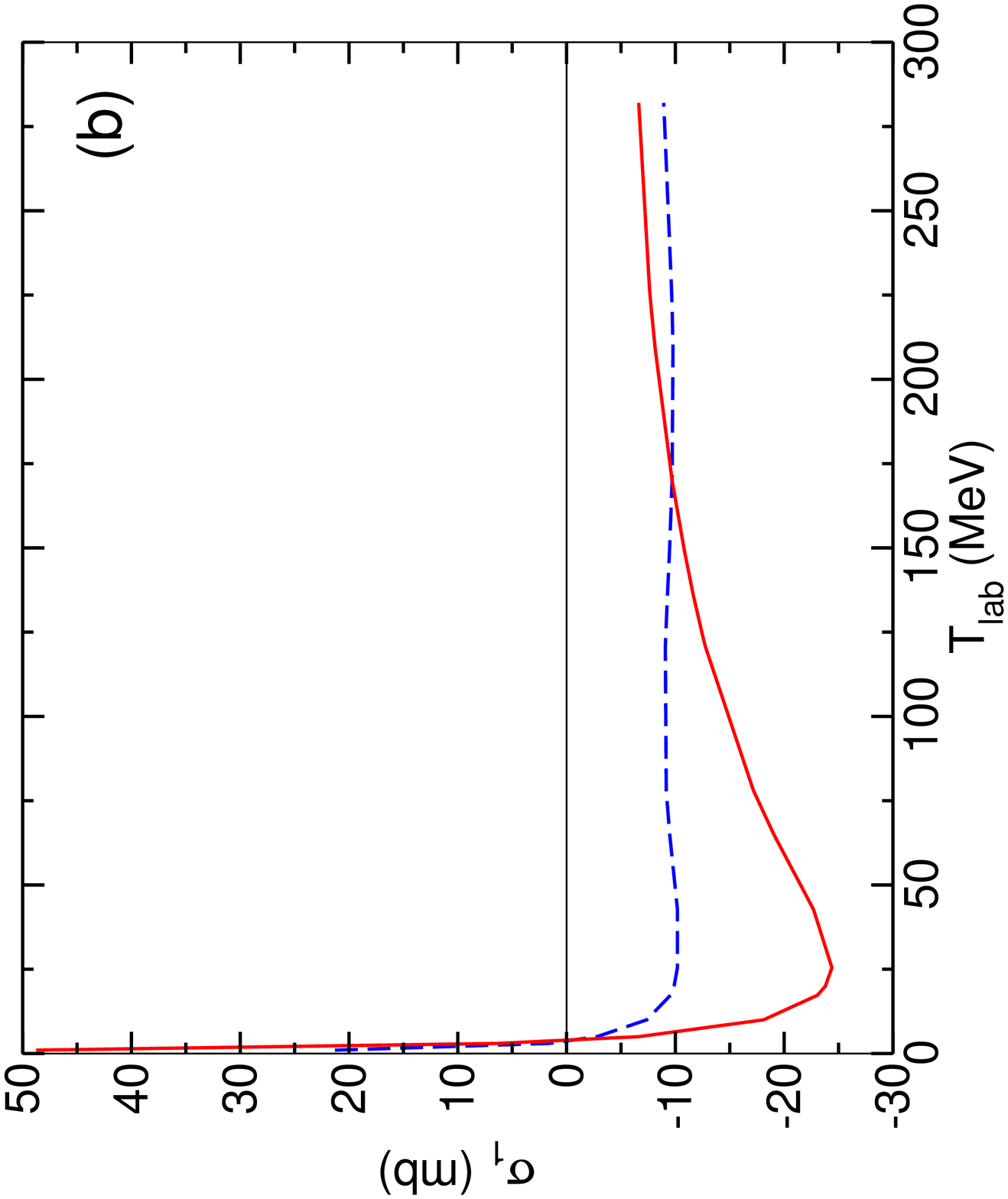}
\includegraphics{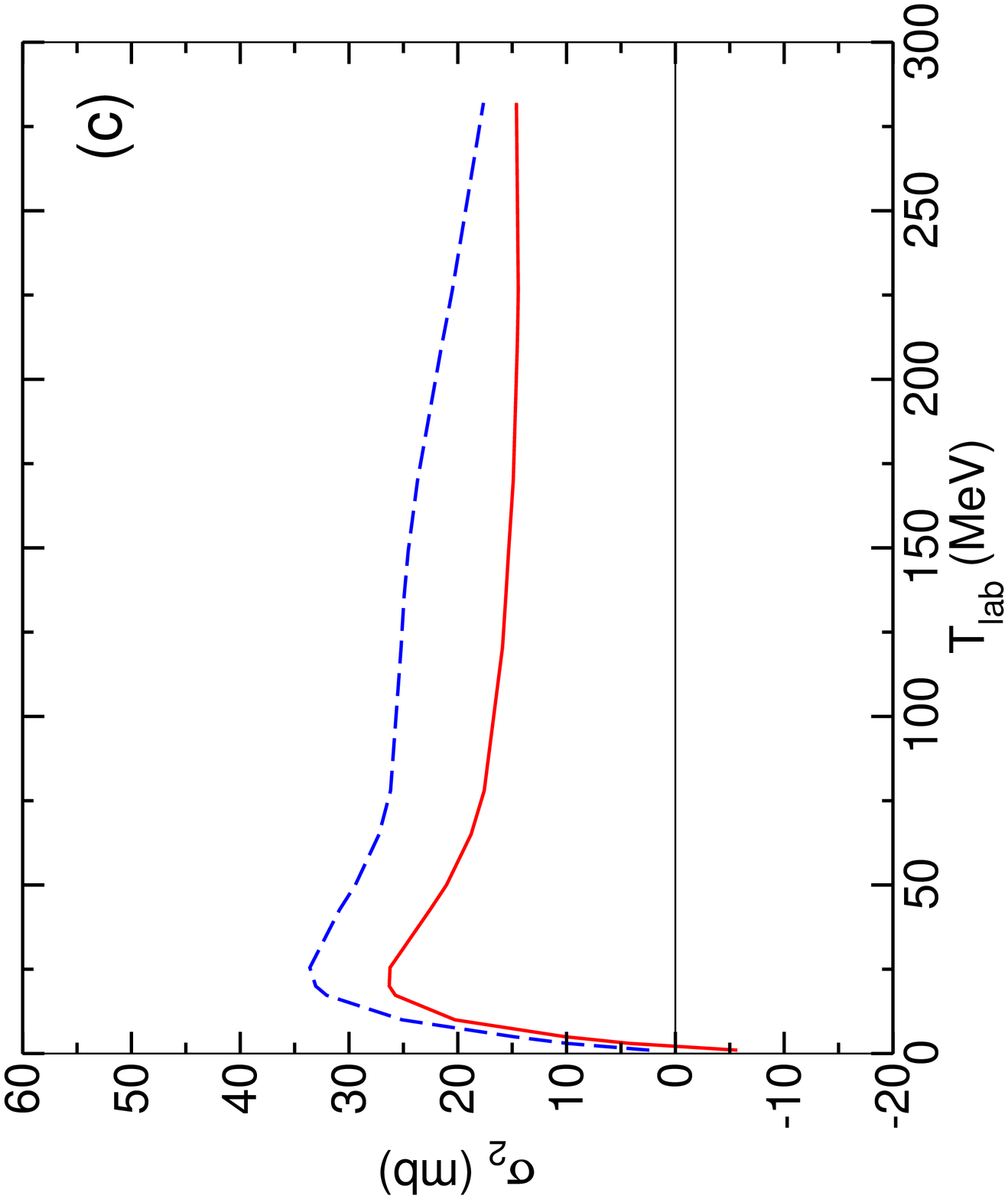}
\vskip 10.5cm
\caption{Total $\bar p n$ cross sections $\sigma_{0}$ (a), 
$\sigma_{1}$ (b), and $\sigma_{2}$ (c)
versus the antiproton laboratory energy $T_{lab}$. 
Results for model D (solid line) and model A (dashed line) 
are shown.
}
\label{fpnJA}
\end{figure}

Let us now come to the spin-dependent $\bar pp$ and 
$\bar p n$ cross sections predicted by the J\"ulich $\bar NN$ interactions.
Corresponding results are presented in Figs.~\ref{totppDA} and 
\ref{fpnJA}. We display the cross sections based on the purely hadronic
amplitude ($\sigma_i^{h}$) and the Coulomb-nuclear interference term
($\sigma_i^{int}$) separately so that one can see the magnitude of the
latter. The total cross sections are then the sum of those two contributions. 
In the concrete calculations the acceptance angle was chosen to be
$\theta_{acc} = 8.8$ mrad \cite{FILTEX}.

At low energies, i.e. around $T_{lab} = 5\sim 10$ MeV, the 
interference terms are comparable to the corresponding purely hadronic 
cross sections and their magnitude increases further
with decreasing energy due to the $1/ k_{\bar p p}$ factor, cf.
Eqs.~(\ref{CNpp2}).
With increasing energy the relevance of the Coulomb-nuclear interference 
terms diminishes more and more in case of the cross sections $\sigma_0$ 
and $\sigma_2$. 
But for $\sigma_1$ the term is still significant, as one can see from 
Fig.~\ref{totppDA}b.
The large magnitude of $\sigma_1^{int}$ as compared to $\sigma_1^{h}$
is due to the fact that $Re M_2^p(0) \gg Im M_2^p(0)$ for 100 - 300 MeV,
for both $\bar NN$ models.
As already pointed out in the context of Eqs.~(\ref{CNpp2}),
$\sigma_1^{int}$ is determined by the former quantity but $\sigma_1^{h}$
by the latter.
Note, that the three cross sections $\sigma_i^{int}$ ($i=0,\dots,2$) 
themselves are all roughly of comparable magnitude for energies from 
around 50 MeV onwards.

While the predictions of the two models for $\sigma_0$ are rather similar
(cf. Figs.~\ref{totppDA}a and \ref{fpnJA}a), even for the Coulomb-nuclear 
interference cross section, this is not the case for the spin-dependent
cross sections $\sigma_1$ and $\sigma_2$. For energies below $T_{lab}
\approx 150$ MeV there are drastic differences between the results based
on the two models. Indeed, for $\sigma_2$ at low energies even the 
sign differs in case of the $\bar pp$ channel. Obviously, here the variations 
in the hadronic amplitude are also reflected in large differences in 
the Coulomb-nuclear interference term. 

For the total $\sigma_1$ (including the hadronic and the Coulomb-nuclear
interference terms) model A predicts a maximum of 12 mb at the beam 
energy $T_{lab}\approx 20$ MeV whereas model D yields a maximum of 
practically the same magnitude at $T_{lab}\approx 10$ MeV. 
In both cases $\sigma_1$ becomes large and
negative at very low energies due to the dominance of the 
Coulomb-nuclear interference term in this region. 
For comparison, in Ref.~\cite{DmitrievMS}, where a version of the Paris 
$\bar NN$ model was employed, the largest value for $\sigma_1$
was found to be -15 mb at $T_{lab}=45$ MeV.
In case of $\bar pn$ scattering both models exhibit a minimum in
$\sigma_1$ at $T_{lab}\approx 25$ MeV and reach values
of around 20 mb (A) and 50 mb (D) close to threshold. 
  
With regard to $\sigma_2$ model A and D predict values around 10 mb  
for $\bar pp$ scattering at higher energies. Close to threshold
large negative values are predicted 
for $\sigma_2^h+\sigma_2^{int}$ due to the Coulomb-nuclear
interference term. One should note, however, that for beam energies below 
5 MeV, say, the total Coulomb cross section becomes very
large. In this case the beam lifetime turns out to be too short and 
the spin-filtering method cannot be used for polarization buildup in 
storage ring.
The results for $\sigma_2$ for $\bar pn$ scattering are comparable
for both models, reaching a maximum of roughly 30 mb around 
$T_{lab}=25$ MeV.

Finally, one should note that for both models A and D 
the polarized cross sections $\sigma_1$ and $\sigma_2$ exhibit
a very different energy dependence in the $\bar pp$ and $\bar pn$
channels. Thus, the expected polarization buildup 
for $\bar p d$ scattering is likewise different from that of 
the $\bar p p$ reaction, as will be shown 
in the next section.
 
%
 
 \section{Results and discussion of $\bar p d$ scattering}

In this section we present numerical results for $\bar p d$ scattering
employing the J\"ulich models A and D \cite{Hippchen,Mull} for the elementary 
$\bar p N$ interactions. 
In order to estimate the role of the double-scattering mechanism, which  
will be not taken into account in our calculation
of the polarized total cross sections, we first calculate the (unpolarized)
total $\bar p d$ cross section and also differential cross sections for 
elastic as well as for elastic plus inelastic scattering 
events within the Glauber theory.
This allows us also to compare our results directly with available
$\bar p d$ experiments so that we can check the reliability of the
approach. 
As was shown in detail in 
Refs.~\cite{Kondratyuksb,Dalkarov,mahalabi,Bendis1,Ma,Bendis2,Robson},
in forward elastic scattering of antiprotons off
nuclei the Glauber theory, though in principle a high-energy approach, 
works rather well even at fairly low antiproton beam energies
like 50 MeV. 
This has to be compared with proton-nucleus scattering
where the Glauber theory is known to give reliable results only for
energies in the order of 1 GeV or above. 
 The Glauber theory is applicable at such low energies, because of the 
 presence of annihilation channels in the $\bar p N $ interaction. 
 Due to strong annihilation effects, specifically in the $S$-waves, higher partial
 waves start to play an important role already fairly close to threshold.
 As a consequence, the $\bar p N$ elastic differential cross section 
 is peaked in forward direction already at rather low energies 
\cite{Mull,Klempt} and, therefore, suitable for application of the
eikonal approximation, which is the basis of the Glauber theory.
In other words it can be seen from the optical theorem that the higher 
the annihilation cross section is, the larger is the modulus of the 
forward elastic scattering amplitude. Indeed, the elastic spin-averaged
$\bar p N$ scattering amplitude can be very well parameterized by 
\begin{equation}
\label{fpn}
  f_{\bar pN}(q)=\frac{k_{\bar pN}\sigma^{\bar p N}_{tot}(i+\alpha_{\bar p N})}
{4\pi} \exp{(-\beta^2_{\bar p N}q^2/2)},
 \end{equation}
 where $\sigma^{\bar p N}_{tot}$ is the total unpolarized $\bar p N$
 cross section, $\alpha_{\bar p N}$ is the ratio of the real to 
 imaginary part of the forward scattering amplitude $f_{\bar pN}(0)$,
 $\beta_{\bar p N}^2$ is the slope of the diffraction cone, $q$ is the
 transferred 3-momentum, and $k_{\bar p N}$ is the $\bar p N$ cms momentum.
 
\begin{table}[t]
\caption{
Parameters of the $\bar p p$ and $\bar p n$ amplitudes according to 
Eq.~(\ref{fpn}) representing the results of the $\bar NN$ models 
A and D of the J\"ulich group \cite{Mull} at different beam energies.
Note that the values for $\sigma_{tot}^{\bar p N}$ and $\alpha_{\bar p N}$
follow directly from the corresponding models, while $\beta_{\bar p N}^2$
is determined in a fit to the corresponding amplitudes.
}
\vskip 0.3cm 
\label{tab1}
\begin{tabular}{|c|c|c|c|c|c|c|c|}
\hline
$T_{lab}$ & $\bar p N$ model & $\sigma_{tot}^{\bar pp}$ & $\beta_{\bar p p}^2$
&$\alpha_{\bar p p}$ &  $\sigma_{tot}^{\bar p n}$ & $\beta_{\bar p n}^2$
&$\alpha_{\bar pn}$ \\ 
MeV & & mb & (GeV/c)$^{-2}$ & & mb & (GeV/c)$^{-2}$ & \\
\hline
10 & A & 409 & 46.24& -0.351 & 372 & 41.1 & -0.372\\
   & D & 441 & 59.1 & -0.20 & 354. & 51.4 & -0.164\\
\hline
25.5 & A & 305 & 46.24 & -0.176 & 267.2 & 36.0 & -0.146 \\
     & D & 312 & 52.4  & -0.130 & 260   &  48.8&  -0.076\\
\hline
50 & A & 233.5&  33   & -0.095 &  209.6 & 26 & -0.034\\
   & D & 240  & 33.47 & -0.121 &  219 & 33.47 & -0.058\\
\hline
78 & A & 209.5  & 29.9  & -0.03 & 192.6 & 26&  0.0417 \\
   & D & 203    &   29.9& -0.03 & 192   & 29.2 & 0.0145 \\
\hline
109 & A & 186 & 25.2 & 0.033 & 171.3 & 24 & 0.108\\
   & D & 175.8 & 25.9 & -0.029 & 165 & 25 & 0.0245\\
\hline
124 & A & 174.4 & 24.45 & 0.057 & 163.   &  24.3 & 0.133\\
    & D & 168.4 & 24.45 & -0.030 & 159.5 & 24.3 & 0.0245 \\
\hline
138 & A &  174.0 &  23.24 & -0.030 & 159.5 & 22.9 & -0.03 \\
    & D &  162.5 & 23.24 & -0.030 & 155 & 22.9  &  -0.03 \\
 \hline
149 & A & 164 &  22.58 &  0.009 & 156.2 & 22.58 & 0.166 \\
    & D & 159.26   & 22.58  & 0.002 & 149 & 22.58&  0.056 \\
\hline
179 & A & 156.3 & 23.4 & 0.110 & 148 & 20.7 & 0.187 \\
    & D & 150. & 23.4 & -0.0256 & 143 & 20.7 & -0.0816 \\
\hline
\end{tabular}
\end{table}

\begin{figure}[b]
\begin{minipage}[c]{80mm}
\includegraphics[width=62mm]{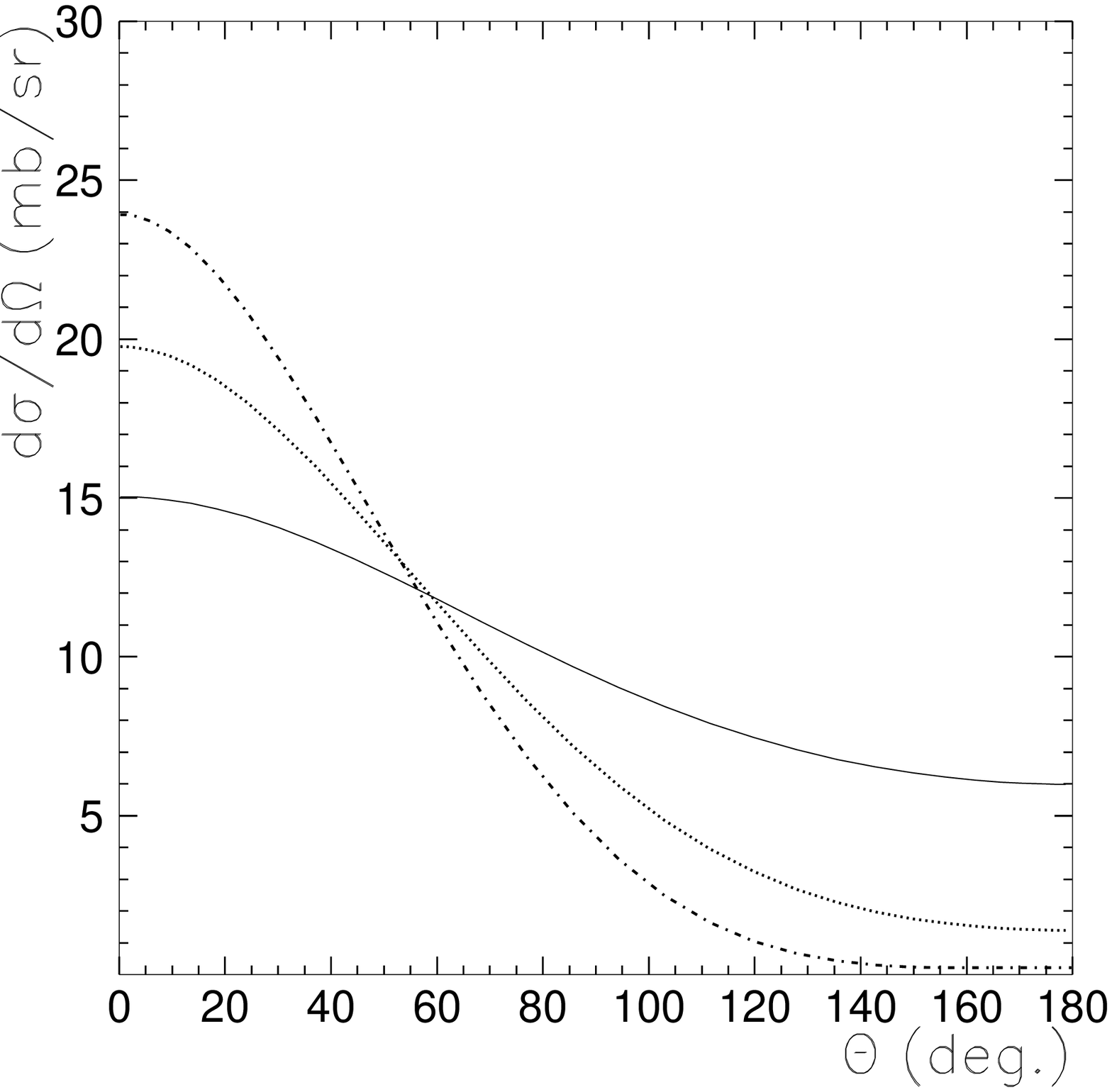}
\end{minipage}
\hspace{\fill}
\begin{minipage}[c]{80mm}
\includegraphics[width=62mm]{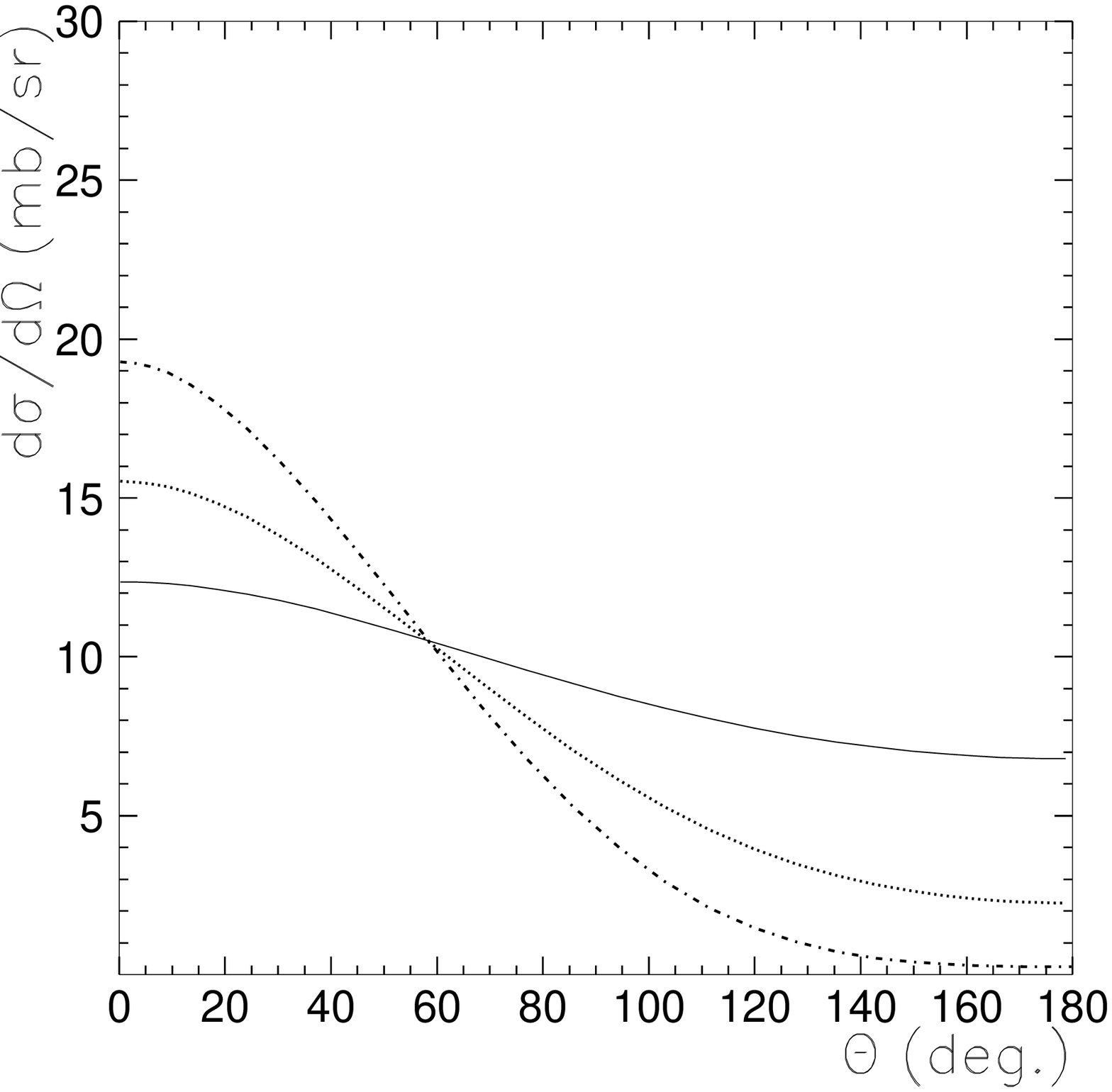}
\end{minipage}
\vspace*{0.5cm}
\caption{Differential cross section of $\bar p p$ (left panel)
and $\bar p n$ (right panel) elastic scattering
predicted by model A
versus the c.m.s. scattering angle $\Theta$ at different beam energies:
50 MeV (dashed-dotted line), 25 MeV (dotted), 10 MeV (solid). 
}
\label{pp179}
\end{figure}

For the present investigation we utilize Eq.~(\ref{fpn}) to represent
the scattering amplitudes of the J\"ulich models in analytical form. 
This allows us then to evaluate the $\bar p d$ scattering amplitude
as given in Eq.~(\ref{Ffi}) in a straight-forward way, including
also the double-scattering correction. 
The values for the parameters $\sigma^{\bar p N}_{tot}$ and 
$\alpha_{\bar p N}$ can be taken directly from the $\bar p N$ scattering 
amplitudes $M_i^N(0)$ that result from the considered J\"ulich models.
The parameters $\beta^2_{\bar p N}$ are determined in a fit  
of $|f_{\bar pN}(q)|^2$ to the unpolarized differential cross section
$d\sigma/d\Omega(\theta)$ resulting from the employed $\bar NN$ models.
We found that even at beam energies as low as 10-25 MeV 
the parameter $\beta^2_{\bar p N}$ is large, i.e. $40 - 50$ (GeV/c)$^2$, 
reflecting the fact that the elastic amplitude in Eq.~(\ref{fpn}) is 
indeed peaked in forward direction. 
For illustration purposes we present $\bar p p$ and $\bar p n$ differential 
cross sections at three selected energies in Fig.~\ref{pp179} for model A.
The concrete parameters for the $\bar pp$ 
and $\bar pn$ amplitudes at the various energies are summarized in 
Table \ref{tab1}. 

\begin{table}
\caption{
Total unpolarized $\bar p d$ cross sections calculated within the 
Glauber theory in the impulse approximation ($\sigma^{IA}_{tot}$)
and with $\bar pN$ double scattering included ($\sigma_{tot}$).
The $\bar N N$ model D of the J\"ulich group \cite{Mull} is used.
Experimental values ($\sigma^{exp}_{tot}$) at the various beam
energies $T_{lab}$ are taken from 
Refs.~\cite{Kalogeropoulos} (a), \cite{Burrows} (b)
and \cite{BizzarriNC74} (c). 
}
\vskip 0.3cm 
\label{tab2}
\begin{tabular}{|c|c|c|c|c|}
\hline
$T_{lab}$  & $\sigma^{exp}_{tot}$  & $\sigma_{tot}$ & $\sigma^{IA}_{tot}$ &
 $R=\sigma^{IA}_{tot}/\sigma_{tot}$ \\
 MeV & mb  & mb &  mb & \\
\hline
57.0 & 415$\pm  16^{a)}$ & 401 & 471& 1.17 \\
58.0 & 390$\pm  15^{b)}$ & 401 & 471& 1.17 \\
57.4$\pm 13.2$ & 366.2$\pm 11.5^{c)}$ & & &\\
\hline
79.8$\pm 10.0$ & 346.4$\pm 8.4^{c)}$ &339 &395 & 1.16 \\
\hline
109.3 & 322 $\pm 16^{a)}$ & 296  & 341& 1.15  \\
109.3$\pm 8.8$ & 310.7$\pm 7.3^{c)}$ & & &\\
\hline
124.1$\pm 8.1$ & 287.5$\pm 6.1^{c)}$ & 286 &328 &1.15\\
126.8 & 295$\pm 7^{b)}$  &    &  &  \\
\hline
137.7 $\pm 7.5$ & 283.4$\pm 5.4^{c)}$ & 277& 317 & 1.15  \\
\hline
147 & 271$\pm 6^{b)}$ & 270  & 308 & 1.14 \\
146.6$\pm 7.1$ & 282.6$\pm 5.4^{c)}$ & & &\\
\hline
179.3 & 269$\pm  7^{b)}$ & 258  &  293 & 1.14 \\
170.5$\pm 8.8$ & 271.8$\pm 4.2^{c)}$ & 264 &301 & 1.14 \\
\hline
\end{tabular}
\end{table}

\begin{figure}[b]
\mbox{\epsfig{figure=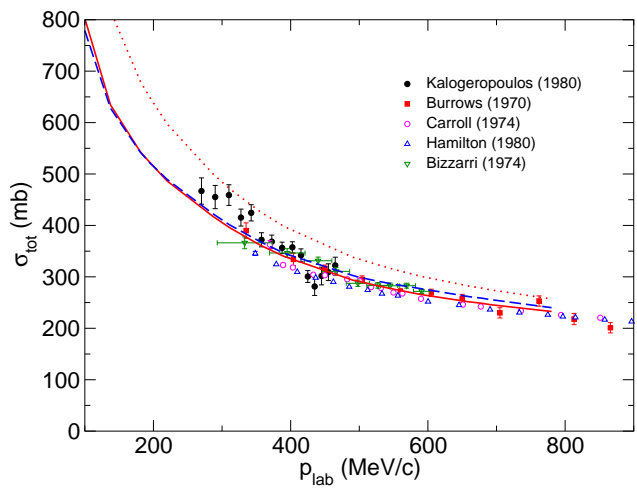,height=0.4\textheight,angle=-90, clip=}}
\caption{Total $\bar p d$ cross section versus the beam momentum 
$p_{lab}$. The solid and dashed lines are results based on the 
$\bar NN$ models D and A, respectively. The dotted line is the 
results for model D obtained within the single-scattering 
approximation. Data are taken from Refs. 
\cite{BizzarriNC74,Kalogeropoulos,Burrows,Carroll,Hamilton}.
}
\label{totpd}
\end{figure}

\begin{figure}
\vspace{12.0cm}
\vglue 0.1cm    
\includegraphics{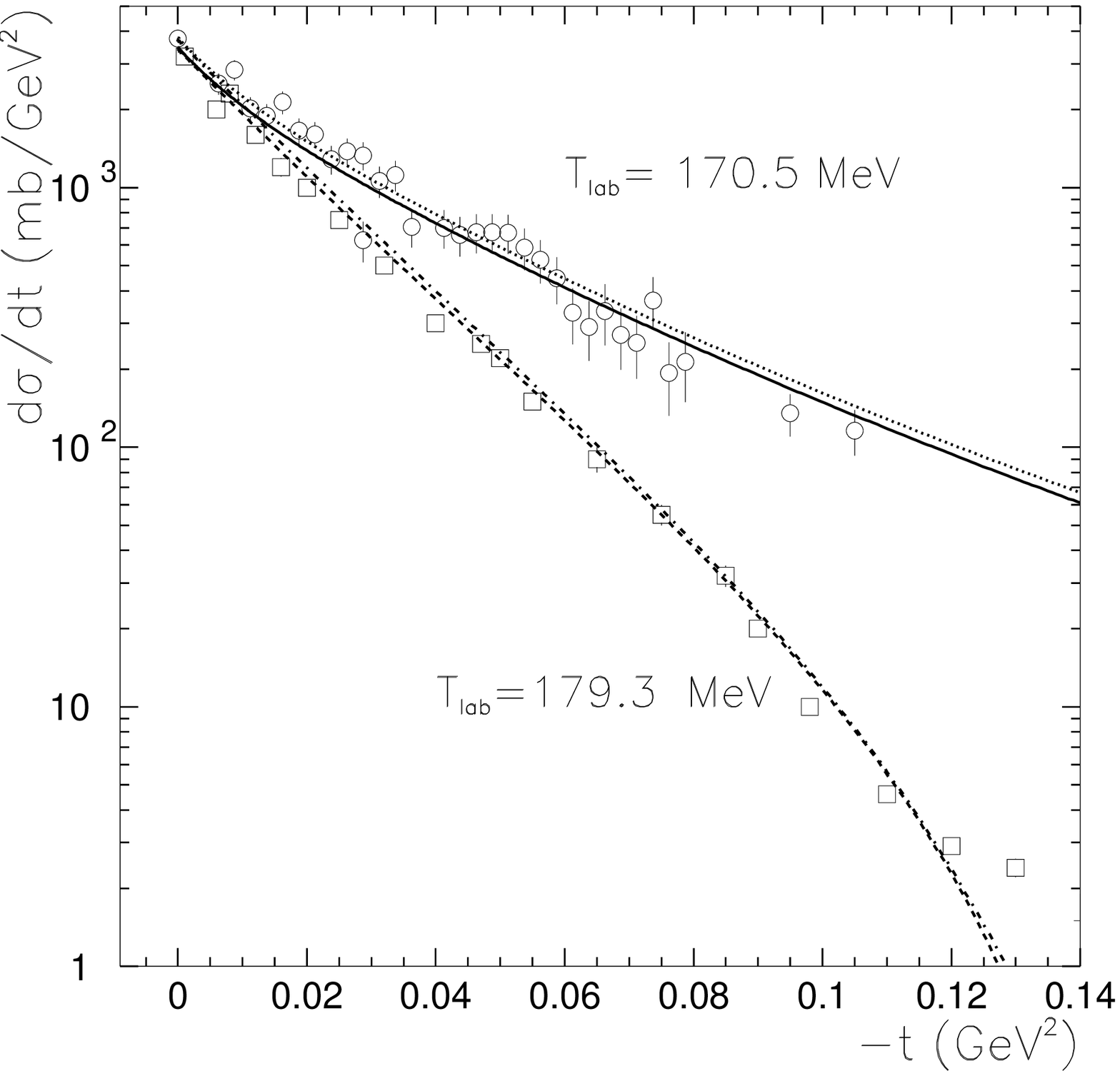}
\includegraphics{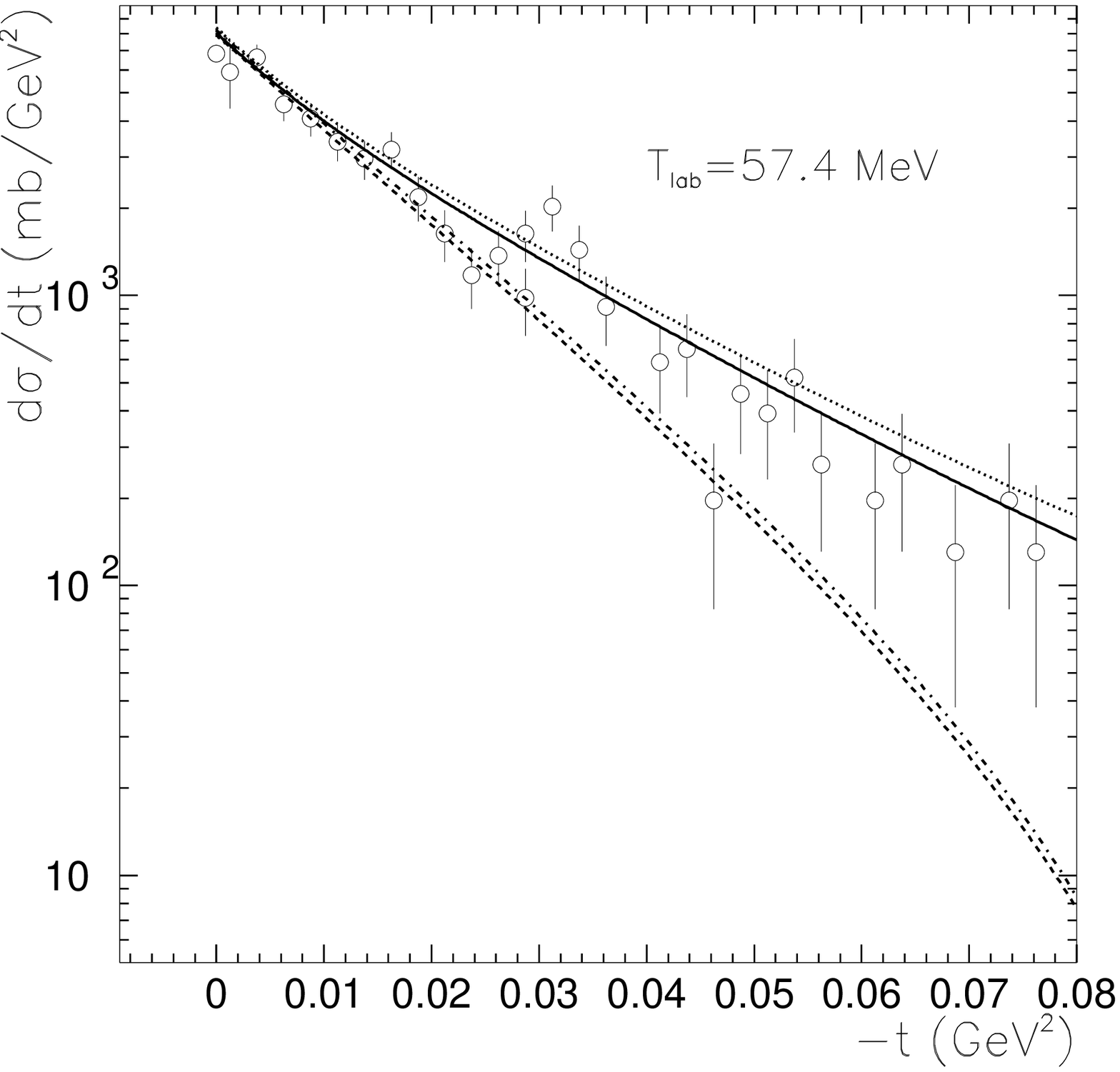}
\includegraphics{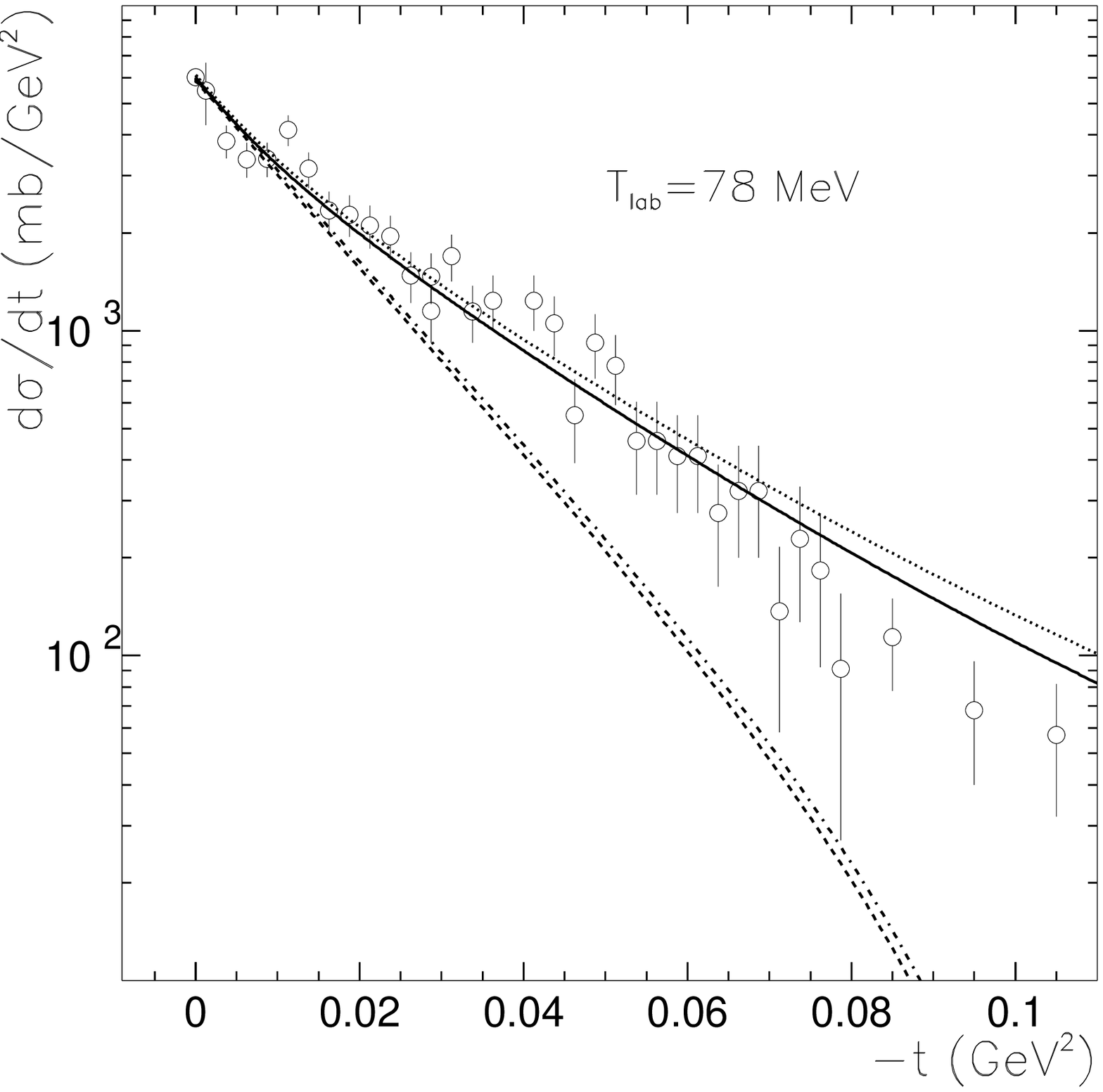}
\includegraphics{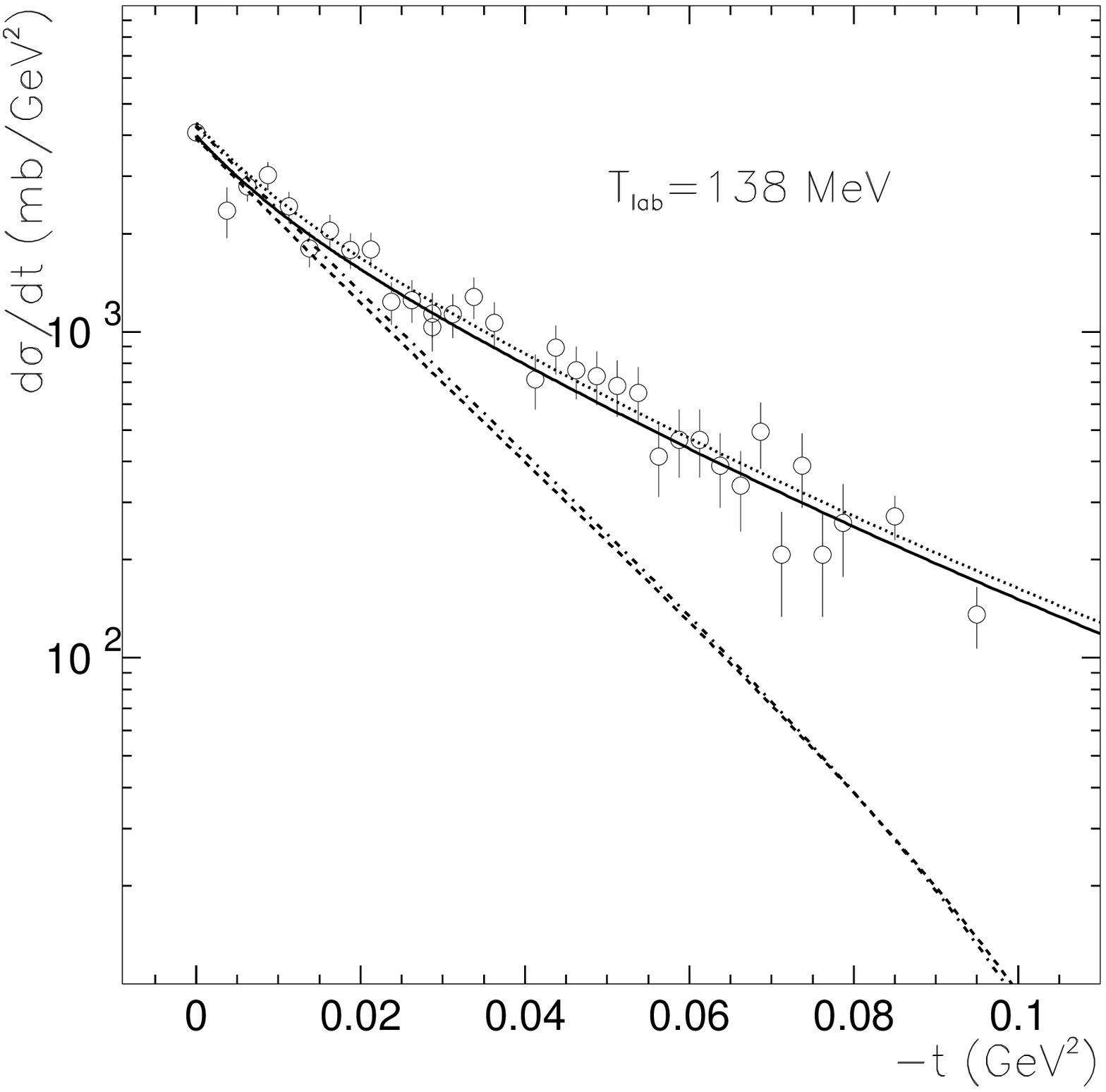}
\vspace{6cm}
\caption{Elastic (lower lines) and elastic plus inelastic (upper lines)
 $\bar p d$ differential cross sections versus the transferred 
 momentum for different $\bar p$ beam energies. 
 The lines are results of a calculation based on the Glauber theory
 for model A (dotted and dashed-dotted) and D (solid and dashed)
 utilizing the parameterizations of the $\bar pN$ amplitudes given in 
 Table \ref{tab1}.
 The ABB form factor \cite{ABB} is used for the deuteron.  
 Data for elastic scattering (179.3 MeV) are taken from 
 \protect\cite{bruge88} (squares) and for elastic plus inelastic
 scattering events (57.4 - 170.5 MeV) from 
 \cite{BizzarriNC74} (circles).
}
\label{pbard179}
\end{figure}

 Results for the total unpolarized $\bar p d$ cross 
 section are displayed in Fig.~\ref{totpd} together with experimental 
 information \cite{BizzarriNC74,Kalogeropoulos,Burrows,Carroll,Hamilton}. 
 A comparision of theory with data at selected energies is presented 
 in Table \ref{tab2} for model D.
 One can see that the single-scattering approximation (shown here only 
 for model D) overestimates the total unpolarized cross section by 
 roughly 15\%, cf. the dotted line in Fig.~\ref{totpd}.
 But the shadowing effect generated by the $\bar p N$ double-scattering 
 mechanism reduces the cross section by about that amount so that
 the final result (solid line) is in good agreement with the experiment. 
 The results for model A are very similar. Thus, as expected the 
 double-scattering corrections to the total unpolarized cross section 
 turn out to be not very large. Actually, even at energies as low as
 10 - 20 MeV they are at most 20-25\%. 
 Therefore, the Glauber theory seems to work rather well for the 
 $\bar p d$ reaction, even at these fairly low energies. 

Predictions for differential cross sections are presented in
Fig.~\ref{pbard179}. 
In the corresponding calculations of the forward $\bar p d$ 
elastic amplitude the single-scattering mechanism as well as the
double-scattering terms were included. 
The ABB form factor \cite{ABB} is used for the deuteron.  
 At $T_{lab}$ = 179.3 MeV data for the elastic differential 
 cross section are available \cite{bruge88}. These data 
 (squares in Fig.~\ref{pbard179}) are nicely reproduced by
 our model calculation for forward angles. Also the 
 differential cross sections for elastic (${\bar p}d\to {\bar p} d$) 
plus inelastic (${\bar p}d\to {\bar p} pn$) scattering events,
 measured at the neighboring energy $T_{lab}$ = 170 MeV 
\cite{BizzarriNC74} (circles), are well described.
 At lower energies no data on the elastic differential cross
 section are available. But there are 
further angular distributions published in Ref.~\cite{BizzarriNC74}. 
 In Fig.~\ref{pbard179} we show results for selected 
 energies, namely 57.4 MeV, 78 MeV, and 138 MeV. 
 Obviously, our model results are in line with those
 data down to the lowest energy. 

\begin{figure}
%
\includegraphics{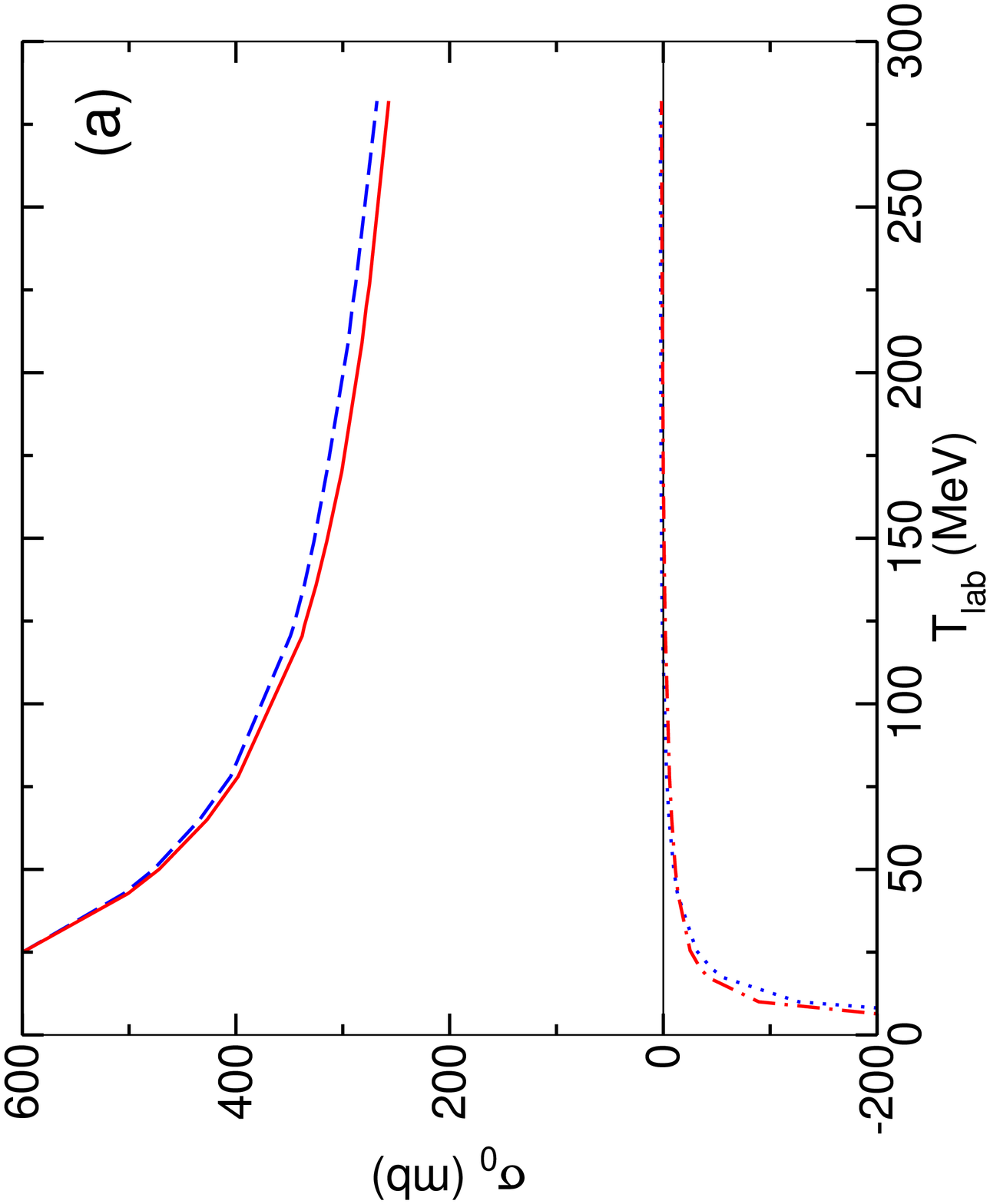}
\includegraphics{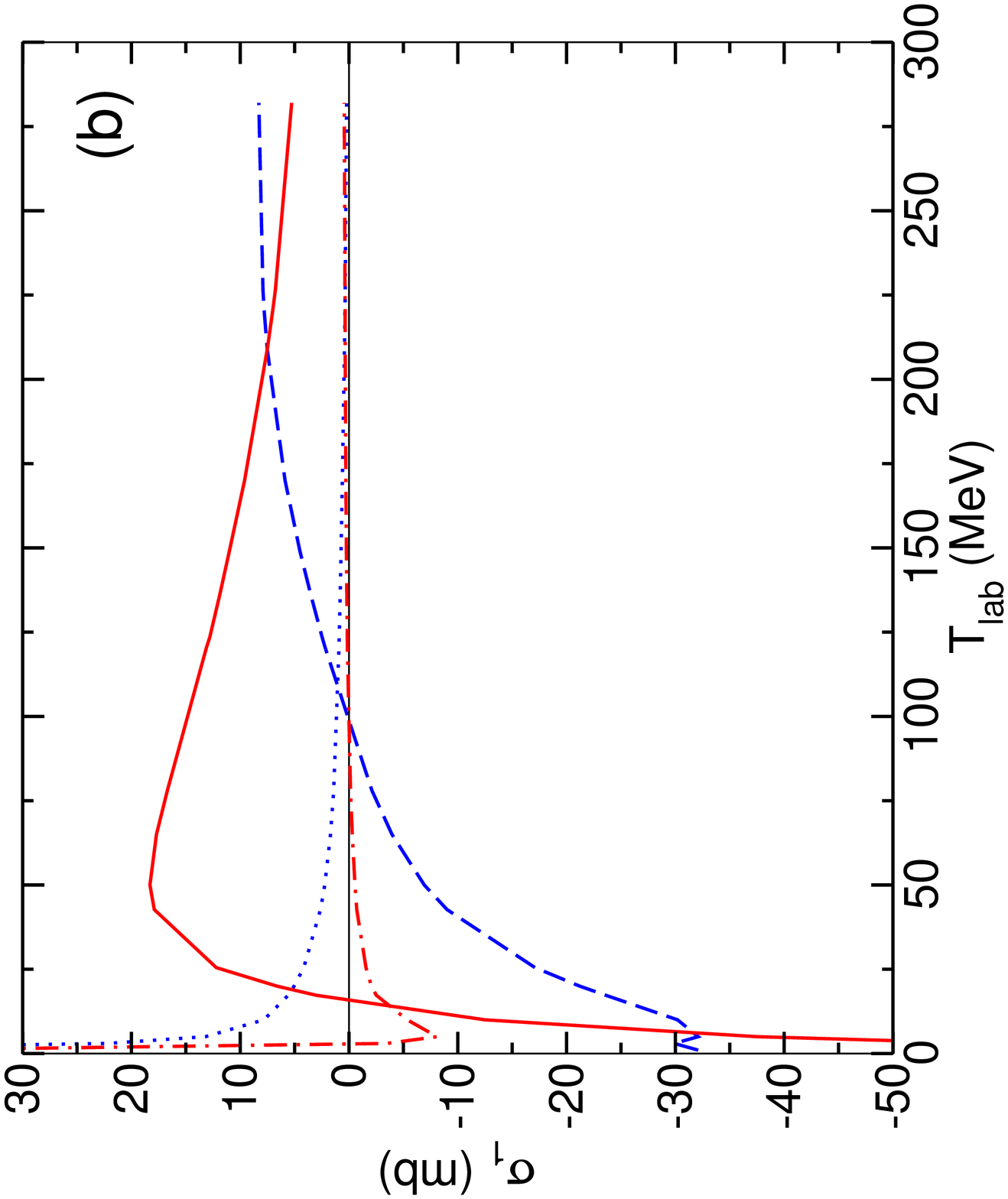}
\includegraphics{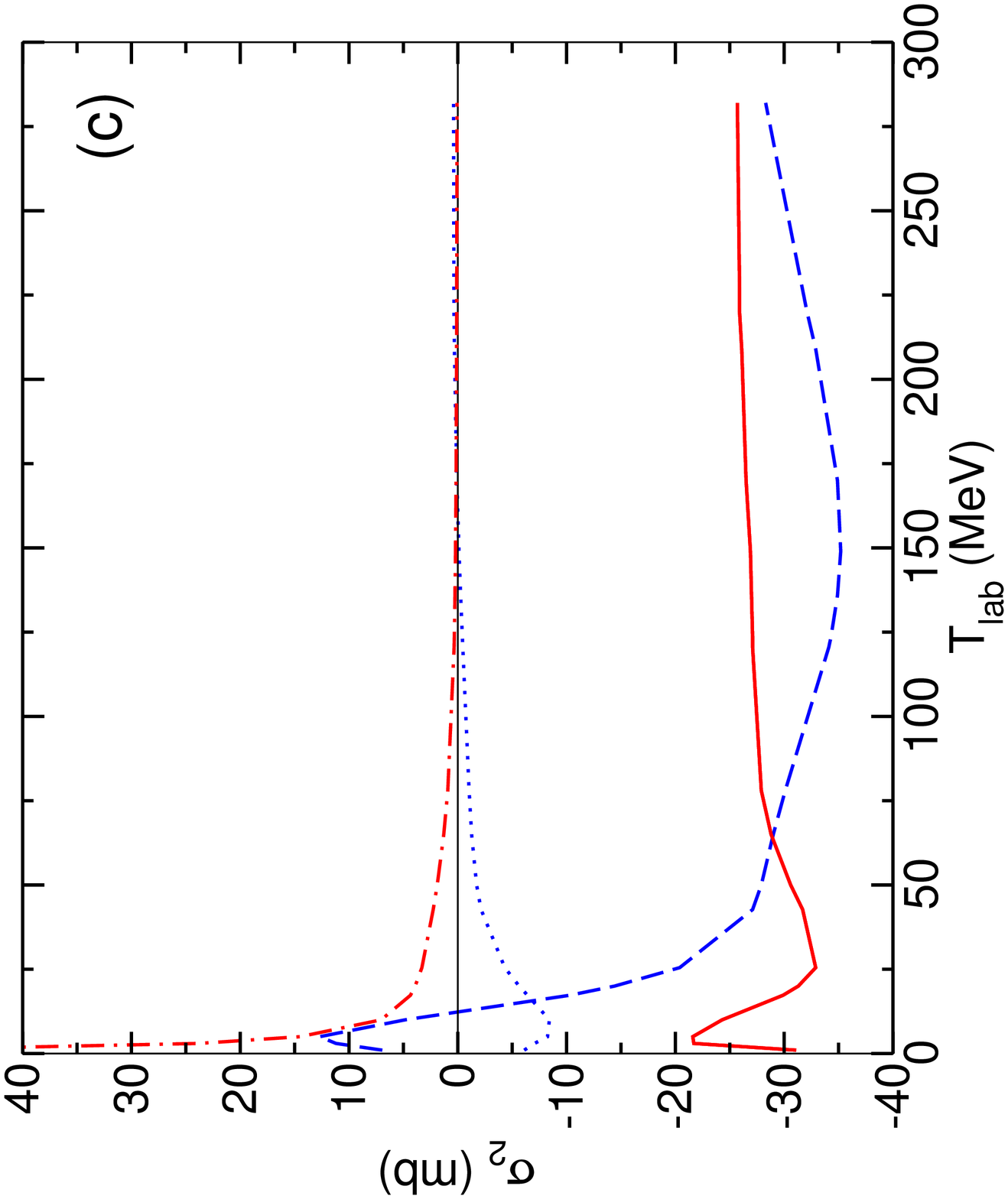}
\vskip 10.5cm
\caption{Total $\bar p d$ cross sections $\sigma_{0}$ (a), 
$\sigma_{1}$ (b), and $\sigma_{2}$ (c) 
versus the antiproton laboratory energy $T_{lab}$. 
Results based on the purely hadronic amplitude,  
$\sigma^h_i$, (model D: solid line, model A: dashed line) 
and for the Coulomb-nuclear interference term, 
$\sigma^{int}_i$, (D: dash-dotted line, A: dotted line) 
are shown.
}
\label{totpdx}
\end{figure}

Results for spin-dependent cross sections are presented in 
Fig.~\ref{totpdx}. Note that here the corresponding calculations
are all done in the single-scattering approximation only. 
In principle, the double-scattering corrections to the 
spin-dependent cross sections could be worked out
by adopting the formalism described in Refs.~\cite{Alberi,Alberi1,Sorensen}. 
But we expect that the double-scattering effects on those quantities 
are roughly of the same magnitude (i.e. less than 20 \% for energies above 
20 MeV) as for the spin-independent cross sections. At least this is what 
we find numerically
within the approximation outlined in Ref.~\cite{Alberi}. 
Therefore, we believe that the single-scattering approximation 
provides a reasonable estimation for the magnitude of the 
polarization buildup effect in $\bar p d$ scattering and we refrain from a
thorough evaluation of the involved double-scattering effects in 
the present analysis.
After all one has to keep in mind that the differences between the $\bar NN$
models A and D introduce significantly larger variations in the cross 
sections $\sigma_1$ and $\sigma_2$, cf. Fig.~\ref{totpdx}b and c.
  
\begin{figure}
\mbox{\epsfig{figure=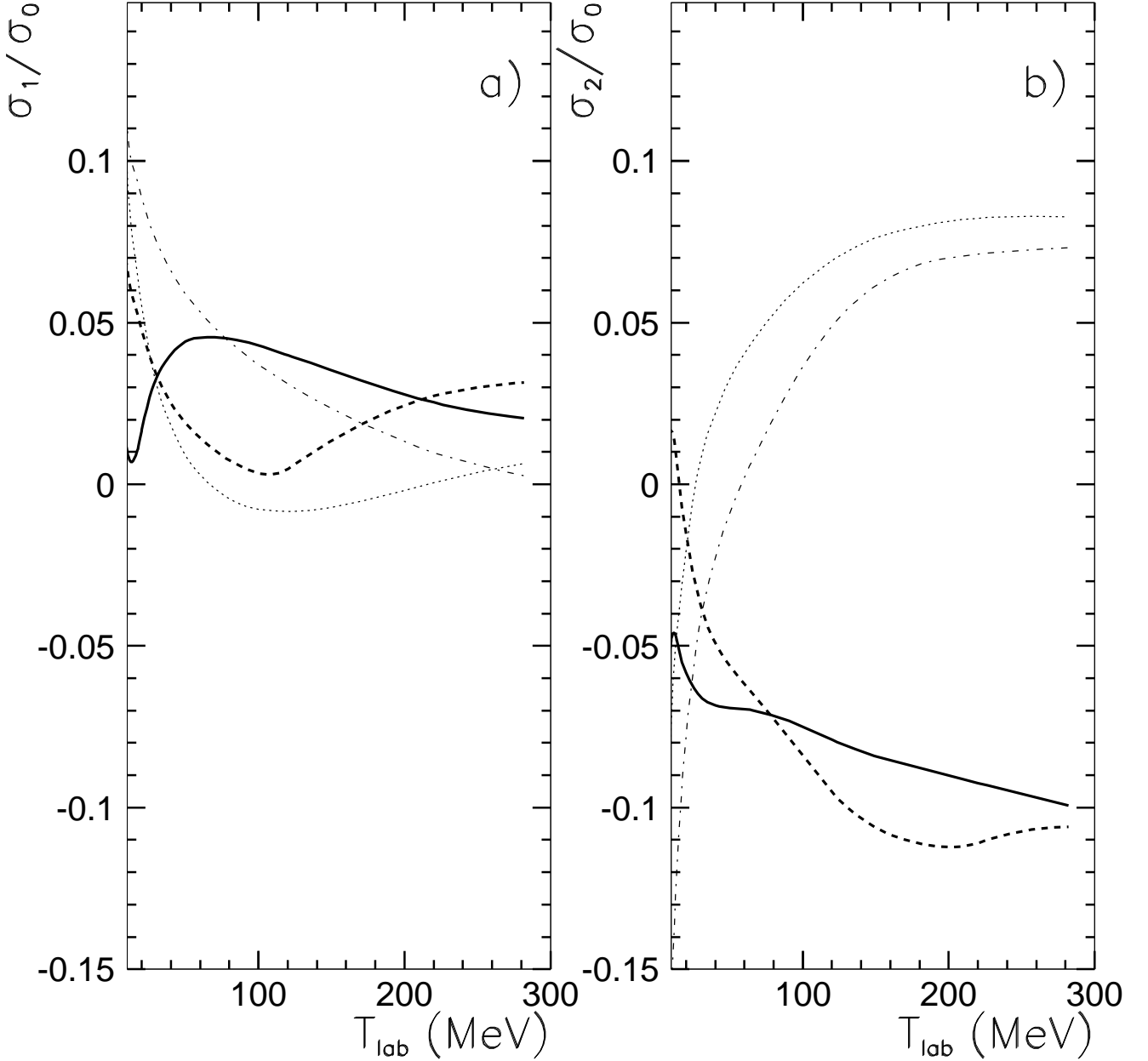,height=0.8\textheight, clip=}}
\caption{Ratio of the total polarized cross sections 
to the total unpolarized cross section,  
$\sigma_1/\sigma_0$ (a) and $\sigma_2/\sigma_0$ (b),
versus beam energy. Results for $\bar p d$ scattering 
[model A (dashed line), D (solid line)] as well as 
for $\bar p p$ scattering 
[model A (dashed-dotted), D (dotted)] are displayed.
The Coulomb-nuclear interference cross sections $\sigma_i^{int}$ $(i=0,1,2)$
for $\theta_{acc}=8$ mrad are included. 
}
\label{ratio2}
\end{figure}

The $\bar NN$ model D predicts large spin-dependent cross section 
$\sigma_1$ with a maximum around 40 MeV. The cross section $\sigma_2$ 
is large as well and practically constant from around 75 MeV onwards. 
In case of model A the maximum for $\sigma_1$ occurs at 
considerably higher energies. Large negative values of $\sigma_1$
 are predicted for energies below 50 MeV with a maximum at 150 MeV.
The predictions for $\sigma_2$ 
are comparable to those of model D for energies above 25 MeV, say. 
As one can see from Fig.~\ref{totpdx}, the largest values for the polarized 
$\bar p d$ cross sections are expected at very low energies,
i.e. for $T_{lab}$ less than 10 MeV, since the $\bar p N$ cross
sections increase with decreasing beam energy.  
However, as was already noted above, at these energies the pure Coulomb cross section
becomes too large, so that the method of spin-filtering for the polarization buildup 
cannot be applied due to the decrease of the beam lifetime.
In any case, while our results at higher energies are expected to be correct within
10\% to 20 \%, there are larger uncertainties below 20 MeV and here our
results should be considered only as a qualitative estimate.
Finally, one can see from Fig.~\ref{totpdx}, that the Coulomb-nuclear interference 
effects become only important for energies below
50 MeV. In contrast to the $\bar pp$ case, for $\bar p d$
$\sigma_1^{int}$ is much smaller than $\sigma_1^h$ at $T_{lab} \ge 100$ MeV.
This different behavior for $\bar p p$ and $\bar p d$ is due to 
the additional terms $M_i^n(0)$ entering the expression for 
$\sigma_i^{int}$ of the latter reaction, as explained in Sect.~\ref{pditerf}. 

The polarization buildup is determined mainly by the ratio of the polarized
total cross section $\sigma_i$ (i=1,2) to the unpolarized one ($\sigma_0$).
Those ratios $\sigma_i/\sigma_0$ are shown in Fig.~\ref{ratio2} for 
beam energies 10-300 MeV for the $\bar p p$ and $\bar p d$ reactions.
The Coulomb-nuclear interference effects are taken into account. 
Once again those results exhibit a significant model dependence. 
However, for all considered cases large values for the ratio 
$\sigma_2/\sigma_0$ of around 10\% are predicted at the higher
energies, which would be sufficient for the requirements of the 
PAX experiment \cite{Frankpc}.
Also, when comparing the $\bar p p$ and $\bar p d$ results 
we see that (the moduli of) the predicted values for the ratio 
$\sigma_2/\sigma_0$ are larger for $\bar p d$ than for $\bar p p$,
for energies of $T_{lab}\approx $ 100 MeV or higher, in case of 
model A as well as for model D. 
Thus, our results suggest that there could be indeed a 
slightly higher efficiency 
for the polarization buildup when using $\bar p d$ instead of $\bar p p$.

\section{Summary}

In the present paper we employed
 two $\bar NN$ potential models developed by the J\"ulich group
 for a calculation of unpolarized $\bar p d$ scattering within the 
 Glauber-Sitenko theory and found that this approach allows one to 
 describe the experimental information on differential and total 
 $\bar p d$ cross sections, available at $T_{lab} = 50-180$ MeV, quantitatively.
 For those spin-independent observables the difference in the 
 predictions based on those two models turned out to be rather small.
The double-scattering corrections to the unpolarized cross section
were found to be in the order of 15\% in the energy range where the
data are available. But we found that even at such low energies
as 10-25 MeV they are not larger than 20-25\%.
This means that, most likely, the Glauber approximation does work 
reasonably well for $\bar p d$ scattering down to fairly small 
energies.  

We also presented results for polarized $\bar p d$ cross sections 
obtained within the single-scattering approximation.
In comparison to $\bar p p$ scattering, for $\bar p d$
scattering there are four total polarized cross sections instead 
of the three in the $\bar p p$ case.
The additional cross section, connected with the tensor polarization 
$P_{zz}^d$ of the deuteron, has no direct influence on the polarization 
buildup, but, in principle, can be used to increase the beam lifetime
by a proper choice of the sign of $P_{zz}^d$.
 
 In the single-scattering approximation the spin dependent $\bar p d$ 
 cross sections are given by the sum of the corresponding $\bar p p$ and 
 $\bar p n$ cross sections. As a consequence, 
 at some energies there is an increase of the polarized cross
 sections as compared to the $\bar p p$ and/or $\bar p n$ case. 
 The Coulomb-nuclear interference effects in the polarized
 cross sections, which  play an important role for the $pp$ and $\bar p p$
 systems, are modified in $\bar p d$ scattering due to the additional
 interference with the purely hadronic $\bar p n$ amplitude. 
 
 The predictions for the spin-dependent cross sections for $\bar p d$ scattering,
 presented in this work, exhibit a fairly strong model dependence, which is due 
 to uncertainties in the spin dependence of the elementary 
 $\bar p p$ and $\bar p n$ interactions. 
 Still, our results suggest that $\bar p d$ elastic scattering can be used 
 for the polarization buildup of antiprotons at beam energies of 10-300 MeV 
 with similar and possibly even higher efficiency than $\bar p p$ scattering.

\subsection*{Acknowledgements} 
We acknowledge stimulating discussions with N.N.~Nikolaev
and F.~Rathmann. Furthermore, we are thankful to A.I.~Milstein,
V.M.~Strakhovenko and H. Str\"oher for reading the manuscript and 
useful remarks. This work was 
supported in part by the Heisenberg-Landau program.

\section*{Appendix: Coulomb-nuclear interference cross
 sections for $\bar p d$ elastic scattering}
\setcounter{equation}{0}
\renewcommand{\theequation}{A.\arabic{equation}}
 
 In the coordinate system with the axes $OX|| {\bf l}$, $OY|| {\bf n}$, 
$OZ|| {\bf m}$, where
\begin{eqnarray}
\label{ISO}
{\bf l}=({\bf k}+{\bf k}')/|{\bf k}+{\bf k}'|, \ 
{\bf m}=({\bf k}-{\bf k}')/|{\bf k}-{\bf k}'|, \
{\bf n}= [{\bf k}\times {\bf k}']/|[{\bf k}\times {\bf k}']|,
\end{eqnarray}
one can find for the
operators ${\hat F}_{\alpha\beta}$ defined by Eq.~(\ref{tfi})
the following structure \cite{KLEVSH}
\begin{eqnarray}
\label{f12}
{\hat F}_{xx}&=&\phantom{-}F_1+F_2{\hat \sigma_y},\ \
{\hat F}_{xy}=\phantom{-} F_7{\hat \sigma_z}+ F_8{\hat \sigma_x},\ \
{\hat F}_{xz}=\phantom{-} F_9+F_{10}{\hat \sigma_y},\nonumber \\
{\hat F}_{yx}&=&-F_7{\hat \sigma_z} + F_8{\hat \sigma_y},\ \
{\hat F}_{yy}= F_3+ F_4{\hat \sigma_y},\ \
{\hat F}_{yz}= F_{11}{\hat \sigma_x} +F_{12}{\hat \sigma_z},\nonumber \\
{\hat F}_{zx}&=& -F_9-F_{10}{\hat \sigma_y}, \ \ 
{\hat F}_{zy}= -F_{11}{\hat \sigma_x} +F_{12}{\hat \sigma_z}, \ \
{\hat F}_{zz}= F_5+ F_6{\hat \sigma_y},
\end{eqnarray}
 where the $F_i$'s are complex numbers and $\hat \sigma_j$ $(j=x,y,z)$ are the Pauli
 matrices. 
 Using Eqs. (\ref{f12}) the spin correlation parameters
can be written in the following form:
\begin{eqnarray}
\label{cij}
C_{x,x}&=&\frac{2}{B}\Bigl \{Im(F_9F_8^*+F_{11}F_3^*+F_{11}F_5^*)
+ Re (F_7F_{10}^*-F_4F_{12}^*+F_{12}F_6^*)\Bigr \}, \nonumber \\
C_{y,y}&=&\frac{2}{B}\Bigl \{Im(F_1F_{10}^*+F_{2}F_9^*+F_{6}F_9^*+
F_{5}F_{10}^*)-
 Re (F_{11}F_{7}^*+F_{12}F_8^*)\Bigr \}, \nonumber \\
C_{z,z}&=&\frac{2}{B}\Bigl \{Im(F_7F_{1}^*+F_{7}F_3^*+F_{9}F_{12}^*)
+ Re (F_{2}F_{8}^*-F_{8}F_4^* +F_{10}F_{11}^* )\Bigr \}, \nonumber\\
C_{x,z}&=&\frac{2}{B}\Bigl \{Im(F_1F_{8}^*+F_{8}F_3^*+F_{9}F_{11}^*)
+ Re (F_{2}F_{7}^*+F_{7}F_4^* -F_{10}F_{12}^* )\Bigr \},\nonumber \\
C_{z,x}&=&\frac{2}{B}\Bigl \{Im(F_7F_{9}^*+F_{3}F_{12}^*+F_{12}F_{5}^*)
+ Re (F_{8}F_{10}^*-F_{4}F_{11}^* -F_{11}F_{6}^* )\Bigr \} ,  
\end{eqnarray}
where $B$ is given by
\begin{eqnarray}
\label{B}
B = \sum_{i=1}^6|F_i|^2+2\sum_{i=7}^{12}|F_i|^2 .
\nonumber 
\end{eqnarray}
The tensor analyzing powers $A_{ij}$ can be expressed in the
following form (see, for example, \cite{KLEVSH}):
\begin{eqnarray}
\label{aij}
A_{xx}&=&\frac{1}{B}\Bigl\{|F_3|^2+|F_4|^2+|F_5|^2+|F_6|^2 
-(|F_7|^2+|F_8|^2 \nonumber \\ 
&&+|F_9|^2+|F_{10}|^2) -2(|F_1|^2+|F_2|^2-|F_{11}|^2-|F_{12}|^2)
\Bigr \},\nonumber \\
A_{yy}&=&\frac{1}{B}\Bigl\{|F_1|^2+|F_2|^2+|F_5|^2+|F_6|^2 
-(|F_7|^2+|F_8|^2 \nonumber \\ 
&&+|F_{11}|^2+|F_{12}|^2)
-2(|F_3|^2+|F_4|^2-|F_9|^2-|F_{10}|^2\Bigr \},\nonumber \\
A_{zz}&=&-A_{xx}-A_{yy}, \nonumber \\
A_{xz}&=&-\frac{3}{B}Re\Bigl [F_7F_{12}^*+F_9F_5^*+F_{10}F_6^*-
F_2F_{10}^*-F_1F_9^*-F_8F_{11}^* \Bigr ].
\end{eqnarray}
The polarized elastic differential $\bar p d$ cross section is
\begin{eqnarray}
\label{diffsec}
\Bigl (\frac{d\sigma}{d\Omega}\Bigr )_{pol}=
 \Bigl (\frac{d\sigma}{d\Omega}\Bigr )_{0}
\Bigl [1+\frac{3}{2}p_j^{\bar p}p_i^dC_{j,i}+
\frac{1}{3} P_{ij}^d A_{ij}+\dots\Bigr ].
\end{eqnarray}
 Here $\Bigl (\frac{d\sigma}{d\Omega}\Bigr )_0$ is 
the unpolarized differential cross section, which is given as
\begin{eqnarray}
\label{diffsecunp}
\Bigr (\frac{d\sigma}{d\Omega}\Bigr )_0=
 \frac{1}{3}\Bigl [
\sum_{i=1}^6|F_i|^2+2\sum_{i=7}^{12}|F_i|^2 \Bigr ] = \frac{1}{3}B.
\end{eqnarray}
In Eq. (\ref{diffsec})
we spell out explicitly only those terms which give a non-zero 
 contribution to the total
 elastic polarized cross section, while other occuring terms are denoted by 
 dots.
 The total elastic polarized cross section can be found by integration of
 Eq. (\ref{diffsec}) over the scattering angle:
\begin{eqnarray}
\label{totelpol}
\sigma_{pol}^{el}=2\pi \int_{\theta_{acc}}^\pi
\Bigl (\frac{d\sigma}{d\Omega}\Bigr )_{pol}\sin{\theta}d\theta.
\end{eqnarray}

 The Coulomb amplitudes are contained only in the following terms:
 $F_1= F_1^h+F_1^C$, $F_3= F_3^h+F_3^C$,
 and $F_5= F_5^h+F_5^C$, where $F_i^h$ is the purely hadronic part of the
 amplitudes in question. 
 Due to the singularity $\sim \sin^{-2}({\theta/2})$
 of the Coulomb amplitudes
 $F_1^C=F_3^C=F_5^C\equiv F^C$ 
 the main contribution
 to the Coulomb-nuclear interference terms in the cross section Eq. (\ref{diffsec}) 
 comes from forward angles and, therefore, is non-vanishing only for those 
 hadronic amplitudes $F_i$ which are non-zero in forward direction.
In the limit of collinear kinematics when $OZ|| {\bf k}||{\bf k}' $
($\theta = 0^o$),
one can find from Eqs. (\ref{fab}) and (\ref{f12}) \cite{uzepan98}
\begin{eqnarray}
\label{collinearF}
F_1^h=F_3^h=g_1, \ \ F_5^h=g_2,\ \ F_7^h= ig_4, \ \ F_{10}^h=-F_{11}^h=-ig_3,\nonumber \\
F_2^h=F_4^h=F_6^h=F_8^h=F_9^h=F_{12}^h=0 .
\end{eqnarray}
 Based on these relations one obtains via Eqs.~(\ref{cij}), (\ref{aij}):
 \begin{eqnarray}
\label{obscollin}
 C_{z,z}(0)=\frac{2}{B}\bigl \{Im F_7(F_1^*+F_3^*)+Re F_{10}F_{11}^*\bigl \},
 \nonumber  \\
 C_{x,x}(0)=C_{y,y}(0)= \frac{2}{B}Im (F_1+F_5)F_{10}^*,
 \nonumber \\
  A_{xx}(0)=A_{yy}(0)=-\frac{1}{2}A_{zz}(0)=
\frac{1}{B}\bigl \{-|F_1|^2+|F_5|^2+
|F_{11}|^2 -|F_7|^2\bigl \}, \nonumber \\ 
C_{z,x}(0)=C_{x,z}(0)=A_{xz}(0)=0.
\end{eqnarray}
Taking into account axial symmetry, which follows from
Eqs.~(\ref{obscollin}), 
the polarized Coulomb-nuclear interference cross section
can be written as 
 \begin{eqnarray}
\label{totpolcoll}
\sigma_{pol}^{int}=2\pi \int_{\theta_{acc}}^{\pi}\Bigl (\frac{d\sigma}{d\Omega}\Bigr
)_{pol} \Bigl \{1+\frac{3}{2}C_{x,x}{\bf P}^{\bar p} {\bf P}^d+
\frac{3}{2}(C_{z,z}-C_{x,x})P_z^{\bar
    p}{P_z}^d+ \frac{1}{3}(A_{xx} -A_{zz})P_{zz}^d\Bigr \}\sin{\theta}d\theta.
\end{eqnarray}

 Using Eqs.~(\ref{collinearF})-(\ref{obscollin})
 and performing the integration over
the scattering angle in Eq. (\ref{totpolcoll}), one can find finally
\begin{eqnarray}
\label{intsecpd}
\sigma_0^{int}&=&\frac{4\pi}{3}
\int_{\theta_{acc}}^\pi
 Re \Bigl \{ [{F_{1}^h}^*(0)+{F_{3}^h}^*(0)+
 {F_{5}^h}^*(0)]F_1^C  \Bigr \}\sin{\theta} d\theta,\nonumber \\
\sigma_1^{int}&=&2\pi\int_{\theta_{acc}}^\pi
 Im \bigl \{ {F_{10}^h}^*(0)(F_3^C+F_5^C) \Bigr \}\sin{\theta}
 d\theta, \nonumber \\
\sigma_2^{int}&=&-2\pi\int_{\theta_{acc}}^\pi
 Im \Bigl \{ [{F_{7}^h}^*(0) +{F_{10}^h}^*(0)](F_1^C+F_3^C)
 \Bigr \}\sin{\theta} d\theta, \nonumber \\
\sigma_3^{int}&=&\frac{4\pi}{3}
\int_{\theta_{acc}}^\pi
 Re \Bigl \{ [{F_{3}^h}^*(0) -{F_{5}^h}^*(0)]
F_1^C \Bigr \}\sin{\theta} d\theta.
\end{eqnarray}

\end{document}